\documentclass[11pt]{article}      
\pdfoutput=1
\usepackage{amssymb}
\usepackage{amsmath}
\usepackage{amsthm}
\usepackage[]{graphicx}
\usepackage[]{subfigure}
\usepackage{tensor}
\usepackage{float}
\usepackage{color}
\usepackage{cancel}
\usepackage{setspace}
\usepackage{fancyhdr}
\usepackage{framed}
\usepackage{braket}
\usepackage{tikz}
\usepackage{comment}
\usepackage{bbold}
\usepackage[noadjust]{cite}

\usepackage[textwidth=6.5in,top=1.2in,bottom=1.2in]{geometry}

\usepackage{setspace}
\usepackage{simplewick}
\usepackage{float}
\usepackage{mathtools}
\usepackage{blkarray, bigstrut} %
\usepackage[bookmarks,linktocpage, colorlinks=true, plainpages = false, citecolor = red,  linkcolor=blue, urlcolor = magenta, filecolor = blue]{hyperref} 
\usepackage{braket}
\usepackage{mathrsfs}
\usepackage[title]{appendix}
\DeclareMathOperator{\arccosh}{arccosh}
\numberwithin{equation}{section}

\begin{document}

\thispagestyle{empty}
\begin{titlepage}
\nopagebreak

\title{\begin{center}\bf Fresh look at the effects of gravitational tidal forces on a freely-falling quantum particle\end{center}}

\vfill

\author{F.\,Hammad$^{1,2,3}$\footnote{\href{mailto:fhammad@ubishops.ca}{fhammad@ubishops.ca}},\,\,P.\,Sadeghi$^1$\footnote{\href{mailto:psadeghi20@ubishops.ca}{psadeghi20@ubishops.ca}},\, N.\,Fleury$^4$\footnote{\href{mailto:flen2202@usherbrooke.ca}{flen2202@usherbrooke.ca}},\, A.\,Leblanc$^{4}$\footnote{\href{mailto:leba3108@usherbrooke.ca}{leba3108@usherbrooke.ca}}}
\date{ }

\maketitle

\begin{center}
	\vspace{-0.7cm}
	{\it  $\,^1$Department of Physics and Astronomy, Bishop's University, 2600 College Street, Sherbrooke, QC, J1M~1Z7 Canada}\\
	{\it  $\,^2$Physics Department, Champlain 
College-Lennoxville, 2580 College Street, Sherbrooke,\\  
QC, J1M~2K3 Canada}\\
{\it  $\,^3$D\'epartement de Physique, Universit\'e de Montr\'eal, 2900 Boulevard \'Edouard-Montpetit,\\
QC, H3T 1J4 Canada}\\
{\it $\,^4$D\'epartement de Physique, Universit\'e de Sherbrooke, Sherbrooke, QC, J1K~2X9 Canada}

	\end{center}
\vfill
\bigskip

\begin{abstract}
We take a closer and new look at the effects of tidal forces on the free fall of a quantum particle inside a spherically symmetric gravitational field. We derive the corresponding Schr\"odinger equation for the particle by starting from the fully relativistic Klein-Gordon equation in order (i) to briefly discuss the issue of the equivalence principle and (ii) to be able to compare the relativistic terms in the equation to the tidal-force terms. To the second order of the nonrelativistic approximation, the resulting Schr\"odinger equation is that of a simple harmonic oscillator in the horizontal direction and that of an inverted harmonic oscillator in the vertical direction. Two methods are used for solving the equation in the vertical direction. The first method is based on a fixed boundary condition, and yields a discrete-energy spectrum with a wavefunction that is asymptotic to that of a particle in a linear gravitational field. The second method is based on time-varying boundary conditions and yields a quantized-energy spectrum that is decaying in time. Moving on to a freely-falling reference frame, we derive the corresponding time-dependent energy spectrum. The effects of tidal forces yield an expectation value for the Hamiltonian and a relative change in time of a wavepacket's width that are mass-independent. The equivalence principle, which we understand here as the empirical equivalence between gravitation and inertia, is discussed based on these various results. For completeness, we briefly discuss the consequences expected to be obtained for a Bose-Einstein condensate or a superfluid in free fall using the nonlinear Gross-Pitaevskii equation.
\end{abstract}
\end{titlepage}

\maketitle
\section{Introduction}\label{Intro}
The motivation behind studying the behavior of a quantum particle in free fall inside the gravitational field of the Earth is primarily to test the equivalence principle at the quantum level. Of course, the problem of the fate of the equivalence principle in quantum theory has given rise to a large body of literature on the subject. Not only various researchers came to various conclusions concerning the satisfaction or the violation of the principle at the quantum level, but even the definition of the principle itself has received many variants and classifications; whence the different conclusions reached by various authors (see e.g., Refs.\,\cite{Sonego1995,Lammerzahl1996,Lorenza,Davies0,Philosophy,NonEquivalenceEP,AmjP,Nature,WEP2019,Will2018} and the references therein). Therefore, instead of distinguishing here between the various definitions of the principle put forward in the literature, we shall only refer to the principle as stating the equivalence between the gravitational mass and the inertial mass of a particle. The former is what should make the particle react to gravity, whereas the latter is what makes the particle react to an accelerated motion.

Such a definition of the principle is empirical in nature and is usually referred to in the literature as the Newtonian equivalence principle\footnote{See, e.g., Ref.\,\cite{NonEquivalenceEP} for a distinction between the weak equivalence principle and the Newtonian equivalence principle, and for other more sophisticated formulations of the principle.}, as opposed to Einstein's equivalence principle. The latter relies on the former to {\it assume} that the physical effects of uniform acceleration are {\it locally} identical to those of a uniform gravitational field. Given the main features of quantum mechanics ---\! nonlocality, the uncertainty principle and the superposition principle\! --- it is indeed clear that one can hardly take Einstein's equivalence principle literally when dealing with quantum objects without adding extra refinements to the definition \cite{Philosophy}. Studying the free fall of quantum objects is therefore tantamount to testing the equivalence of the two concepts of mass by conducting quantum analogs \cite{Davies,R3,R4,R5,R6,R7,R8,R10,QuantumGalileo,NonCommutativeQM,R2,AntiMatterFF2015,EP2017,R1,Colcelli(2020)} of Galileo's experiment of dropping massive objects. Indeed, the quantum analogs of the latter can perhaps be viewed as testing the concept of mass beside testing the universality of free fall, for in quantum mechanics a mass-independent behaviour can rarely be found \cite{Sonego1995,Philosophy}. Starting with the theoretical study in Refs.\,\cite{QBounce1971,COW1974}, testing the free fall of a quantum particle has, by now, given rise to very well established experimental results \cite{COW1975,COWReview,DroppingAtoms,QBounceNature1}. The latter experiments, and the more recent works on testing the effect of gravity on quantum particles \cite{QBounceNature2,QBouncePhysToday2002,R9,QBounce2003,Landry,FFReview2017,Science,GravityLandauI,GravityLandauII,Josephson,COWBall,QHE}, mainly rely on cold neutrons and neutral atoms. 

The effect of the linear gravitational field near the surface of the Earth on cold neutrons has been demonstrated based on two main approaches. The first approach consists in inducing a quantum interference between two superposed branches of a split wavefunction of a particle, of gravitational mass $m_g$ and of inertial mass $m_i$. In the Colella-Overhauser-Werner (COW) experiment \cite{COW1975}, one branch propagates horizontally at a different height relative to where the other branch is propagating. The branch propagating at height $z_0$ acquires a gravitational potential energy $\Delta U=m_ggz_0$ higher than the one propagating at height $z=0$, where $g$ is the gravitational acceleration at the location of the experiment. As a consequence, after traveling a distance $L$ along the higher path, the particle acquires due to the change $\Delta p$ in its linear momentum an extra phase of $\Delta\phi=\frac{1}{\hbar}\Delta p\, L$ relative to the phase it acquires when traveling along the lower path. For a wavepacket of wavelength $\lambda_0=h/p_0$, the conservation of the total nonrelativistic energy $p^2/2m_i+U$ leads then to
\begin{equation}\label{COWPhase}
\Delta\phi\approx-\frac{m_im_gg}{2\pi\hbar^2}z_0\lambda_0L.
\end{equation}
This phase difference gives rise to a measurable interference between the two branches \cite{COW1975}. The experiment has been done using cold neutrons to benefit from their long wavelength $\lambda_0$. Note, however, that, as we intentionally made it explicit here, what appears in Eq.\,(\ref{COWPhase}) is the product $m_im_g$. This is the reason why such an experiment relying on quantum interference does not yield a mass-independent result even if $m_i=m_g$. Similarly, in the much more recent and precise experiments relying on quantum interference the measured differential phase shift is proportional to $k_{\rm eff}\,g T^2$, where $k_{\rm eff}$ is the effective wave vector of the falling atoms and $T$ is the time between pulses in the light-pulse atom interferometry experiments (See, e.g., the up-to-date comprehensive review given in Ref.\,\cite{R3}). The inertial mass in this case is thus contained inside $k_{\rm eff}$. A more direct quantum analog of the mass-independent free-fall experiments would require an equation involving the {\it isolated} ratio $m_g/m_i$. Unfortunately, this is rarely the case \cite{Sonego1995,NonEquivalenceEP} as the two concepts of mass are rather involved in very subtle ways in quantum dynamics \cite{R9,R10}. Nevertheless, we shall see in this paper that such a case does actually arise when taking into account tidal forces. As we shall discuss it, however, general relativity itself does not allow the appearance of $m_g$ in the equations.

Another method for testing the equivalence of the two masses using cold neutrons consists in inducing a quantum bounce on the quantum particle \cite{QBounceNature1}. As the particle falls downwards and ``hits" the ground and ``bounces" off, the particle acquires a discrete energy, for in the bouncing process the particle ``interferes" with its own wavefunction. The equation used to describe the dynamics of the neutrons inside the gravitational field of the Earth is the Schr\"odinger equation with a linear external interaction potential given by $m_ggz$, where $z$ is the vertical distance measured from the surface of the Earth. The general solution $\psi(z)$ to the resulting time-independent Schr\"odinger equation, with such a positive and linear potential, is a linear combination of the Airy functions ${\rm Ai}(z)$ and ${\rm Bi}(z)$. Discarding the unphysical solution ${\rm Bi}(z)$ ---\! as it is divergent at infinity in $z$\! --- one keeps only the function ${\rm Ai}(z)$. After requiring the boundary condition $\psi(z=0)=0$ to be satisfied by the wavefunction, as imposed by a fixed reflecting horizontal flat surface at $z=0$, the Airy function ${\rm Ai}(z)$ leads to the following approximate expression for the discrete energy of the freely-falling neutrons \cite{QBounce2003}:
\begin{equation}\label{AiryQuantization}
E_n\approx\left[\frac{9m_g^2\pi^2\hbar^2 g^2}{8m_i}\left(n-\frac{1}{4}\right)^2\right]^{\frac{1}{3}},
\end{equation}
where $n$ is a positive integer. Note that, although it was not done so in Ref.\,\cite{QBounce2003}, we have, for later reference, explicitly shown here the appearance of the ratio $m_g^2/m_i$ as opposed to an isolated ratio $m_g/m_i$. Now, physically, the emergence of this quantization condition in the energy spectrum of the particle can be understood as originating from the nonlocality of the wavefunction. In fact, the latter ``feels" simultaneously the lower boundary condition $\psi(0)=0$, imposed on it by the presence of the horizontal flat surface from below, and the ``pushing" down of the gravitational force from above. The effect of gravity on the quantum particle does not seem thus to differ much from the particle-in-a-box effect, which is mainly due to the spatial extension of the wavefunction. 

When we know that real gravity exhibits tidal forces, in contrast to the one due to an accelerated frame, it becomes clear that the spatial extension of the wavefunction has not been fully exploited in such a study. Indeed, the effect of gravitational tidal forces on the freely-falling quantum particle is simply ignored. For works on tidal-force effects on quantum systems, see, e.g.,  Refs.\,\cite{Speliotopoulos,Davies,R3,R4,R5,R6,R7,R8}, as well as Ref.\,\cite{Gabel} where tidal-force terms in curved spacetime have been exposed in detail but their consequences remained unexplored. One of our two goals in the present paper is to contribute to the existing literature on this subject by systematically carrying out a detailed investigation of three new different configurations of a quantum system under the influence of gravitational tidal forces. 

Of course, it is evident that tidal forces can only bring very weak effects on the dynamics of a freely-falling quantum object. Nevertheless, it is very interesting and very important to investigate them more rigorously at the theoretical level as they might very well become, sooner than expected, accessible at the experimental level as well. For, as far as we can tell, only tidal forces allow us to distinguish between gravity and acceleration. As such, investigating the effects of tidal forces promises to shed more light on the (non)equivalence between gravitational effects and inertial effects at the quantum level. The second goal of the present paper is specifically to take into account the effects of tidal forces on the free fall of a quantum particle to come up with new perspectives on the equivalence principle at the quantum level. At first sight, this proposal might seem meaningless and at odds with what we saw above concerning the locality requirement in Einstein's equivalence principle. It turns out, however, that in contrast to focusing on point-like systems, the effect of gravitational tidal forces at the quantum level brings a new life to the distinction ---\! already offered by Newtonian gravity\! --- between the two concepts of mass thanks to the extended nature of the wavefunction.

In fact, a different, and more rigorous, route to arrive at the phase difference $\Delta\phi$ used in the COW experiment, would be to compare the two forms displayed by the Schr\"odinger equation governing the evolution of the wavefunction but written in two different reference frames: the noninertial laboratory frame held fixed to the ground, and an inertial reference frame translating downwards with the uniform acceleration $a=g$. The dynamics of the particle subjected to the linear gravitational field is described in the noninertial-frame coordinates $(t,z)$ by a Schr\"odinger equation displaying the linear potential term $m_ggz\psi$. Going to the freely-falling inertial frame, of acceleration $a$ and of coordinates $(t',z')$, the Galilean change of coordinates $t'=t$ and $z'=z+\frac{1}{2}a{t}^2$ transforms the Schr\"odinger equation into an equation with a time-dependent term and with a term proportional to the gradient of the wavefunction $\psi$.
It was shown in much detail in Ref.\,\cite{AmjP} that in order for the resulting equation to take the form of the Schr\"odinger equation of a free particle described by a wavefunction $\varphi(z',t')$, two conditions must be satisfied. The first, is that the acceleration $a$ of the falling frame should actually be $a=\frac{m_g}{m_i}g$. The second is that the wavefunction $\varphi$ used in the inertial frame differs from the wavefunction $\psi$ used in the noninertial frame by a phase factor only:
\begin{equation}\label{WaveFunctionsLink}
\psi(z',t')=\varphi(z',t')
\exp\left(-\frac{im_ia}{\hbar}z't'+\frac{im_ia^2}{3\hbar}{t'}^3\right).
\end{equation}
To see that this result yields the same phase difference (\ref{COWPhase}) obtained by other means, one only needs to express the phase factor in Eq.\,(\ref{WaveFunctionsLink}) in terms of the coordinates $(t,z)$ of the noninertial frame fixed on the ground. In fact, the phase factor takes then the form, $\exp(-\frac{im_ia}{\hbar}zt-\frac{im_ia^2}{6\hbar}t^3)$ \cite{AmjP}. Recalling that the time the particle spends in the branch at the height $z_0$ is $t=L/v_0$ and that the speed $v_0$ is related to the wavelength by $\lambda_0=h/(m_iv_0)$, we immediately see that the first term inside the parentheses leads indeed to the phase difference (\ref{COWPhase}). Very important here, is the fact that recovering the phase difference (\ref{COWPhase}) is possible only if the acceleration $a$ of the falling frame is chosen such that $a=\frac{m_g}{m_i}g$. 

The second term proportional to $t^3$ in this phase factor does not contribute to the phase difference in such a setup, because it is the same for the two branches as it does not depend on $z$. It seems then that such an extra term could be considered as a mere gauge artifact that arises due to the Galilean change of coordinates between the two reference frames. This is not the case, however. In fact, it was argued in the recent Ref.\,\cite{t3Phase} (see also Refs.\,\cite{R3,R10}) that such a term could be tested experimentally by first splitting the wavefunction into a superposition of two branches on the same horizontal plane and then making, somehow, the particle in one of the two branches drop downwards so that it acquires the above relative phase factor. In the reference frame of the freely-falling particle, which has the acceleration $a$ relative to the other branch, the particle's wavefunction should be perceived as that of a free particle, $\varphi$, whereas relative to the horizontally-moving branch the wavefunction should be perceived to be $\psi$.
Therefore, the term proportional to $t^3$ in the relative phase factor would, in principle, also contribute to the phase difference giving rise to the interference. This would show the equivalence between inertia and gravitation.

In this paper, we show that tidal forces provide an alternative strategy for investigating analogs of such a phase-difference effect. The strategy avoids the delicate need to make the particle in one branch suddenly drop downward to meet the other branch. In addition, the strategy avoids the need to associate a reference frame to the quantum particle, and hence avoids attaching a constant and precise acceleration to a particle that is supposed to be a de-localized object governed by probabilities and which is in a superposition along two different paths. As such, the requirement $a=\frac{m_g}{m_i}g$, that actually hides a distinction between inertia and gravitation in Eq.\,(\ref{WaveFunctionsLink}), is avoided. The trick is, indeed, remarkably played by the hitherto neglected tidal forces. The idea is to associate an inertial frame to a freely-falling laboratory, rather than to the freely-falling quantum particle. The use of Fermi normal coordinates in such a frame is then justified, and one would thus immediately have access to ``pure" gravitational tidal forces, in the sense that the latter are cleaned up from the noninertial contamination experienced by a fixed reference frame on the ground. As we shall see, this leads to completely different and new perspectives than by searching for phase differences of the forms (\ref{COWPhase}) and (\ref{WaveFunctionsLink}). 

The remainder of this paper is structured as follows. In Sec.\,\ref{Sec:II}, we examine in great detail the effect of tidal forces on the free fall of a quantum particle. We first derive the Schr\"odinger equation from the Klein-Gordon equation in order to compare the tidal forces to the non-tidal relativistic corrections due to a curved spacetime. We then use two approaches to solve the Schr\"odinger equation. First, we use a time-independent boundary condition, and then we use time-dependent boundary conditions. In both approaches, the energy spectrum of the particle comes out discrete. However, the important difference is in the specific forms in which the masses $m_i$ and $m_g$ appear in each result. We then examine in Sec.\,\ref{Sec:III} tidal forces in the inertial frame, and show how a substitute for the $t^3$-phase could be investigated, bringing a different perspective on the interplay between inertia and gravitation. In Sec.\,\ref{Sec:III}, we briefly discuss how our work could be extended to include the effects of tidal forces based on the Gross-Pitaevskii equation used to describe Bose-Einstein condensates and superfluids. We conclude this paper with a summary and discussion section.   

\section{Tidal forces in the noninertial frame}\label{Sec:II}
Although our investigation in this section will be based on the nonrelativistic Schr\"odinger equation, it is important to extract the latter from the fully relativistic Klein-Gordon equation in curved spacetime. There are three reasons behind such an enterprise. The first reason is that one can properly take into account the effects of curved spacetime on the wavefunction of the particle only within the Klein-Gordon equation. The second reason is that, being interested in tidal forces which consist of very small perturbations to the linear gravitational potential, one needs to make sure that such terms still dominate compared to the relativistic correction terms before discarding the latter. The third reason is that starting from the Klein-Gordon equation rather than assuming beforehand the Schr\"odinger equation, allows us to neatly see why, in contrast to the latter equation, the former does not allow one to consistently distinguish between a gravitational mass and an inertial mass. Such a consideration is what allows us indeed to discuss the Newtonian equivalence principle in quantum mechanics.  

The Klein-Gordon equation for a particle in a curved spacetime reads $(\Box-\frac{m_i^2c^2}{\hbar^2})\varphi=0$, where the d'Alembert operator is given by\footnote{The signature of the spacetime metric $g_{\mu\nu}$, and of its inverse $g^{\mu\nu}$, is here taken to be $(-,+,+,+)$. Greek indices denote the four spacetime coordinates and take the values $\{0,1,2,3\}$, whereas the Latin indices will stand for the three spatial coordinates and take the values $\{1,2,3\}$.}, $\Box=g^{\mu\nu}\nabla_\mu\nabla_\nu$. When writing explicitly the covariant derivatives $\nabla_\mu$ in terms of the Christoffel symbols $\Gamma_{\mu\nu}^\lambda$, the Klein-Gordon equation takes the form,
\begin{equation}\label{KGCST}
\left(g^{\mu\nu}\partial_\mu\partial_\nu-g^{\mu\nu}\Gamma_{\mu\nu}^\lambda\partial_\lambda-\frac{m_i^2c^2}{\hbar^2}\right)\varphi=0.
\end{equation}
The preliminary remark in relation to the Newtonian equivalence principle that should be made here is that the gravitational interaction manifests itself in this equation only through the partial derivative operators $\partial_\mu$ acting on the wavefunction. As such, no distinction can be made at this point between a gravitational mass $m_g$ (which is completely absent here) and the inertial mass $m_i$ of the particle. In other words, one can say that, in some sense, in the curved-spacetime Klein-Gordon equation, gravity does not ``act" on the particle but on the wavefunction of the latter. As the wavefunction is shaped by the inertial mass-energy of the particle, gravity has no other option but to act on the particle through the  inertial properties of the latter. We shall come back to this observation in the conclusion section.

In fact, in order to display explicitly the coupling of the mass of the particle to the gravitational field, we need to adopt, as is customary, the ansatz $\varphi(x)=\psi(x)e^{-\frac{i}{\hbar}m_ic^2t}$ for the wavefunction. The mass $m_i$ appearing in the exponential in this ansatz is not a gravitational mass, but rather the inertial mass. This is because the special relativistic rest-mass energy $m_ic^2$ is tied up to the inertial mass, not to a gravitational one. Thus, in contrast to the Schr\"odinger equation, the equivalence of the two masses is already a built-in assumption in the Klein-Gordon equation in curved spacetime. This, in turn, is hardly surprising as it is tied up to the way general relativity is founded. Therefore, according to Eq.\,(\ref{KGCST}), the only possibility to allow for the existence of a mass $m_g$ different from $m_i$ would be to artificially attach the ratio $m_g/m_i$ to the metric itself. However, this would greatly alter the main design of general relativity, especially as it would require a different metric for different test particles in case the ratio $m_g/m_i$ is different for different particles.

Let us now insert that ansatz for $\varphi(x)$ into Eq.\,(\ref{KGCST}) and consider a spherically symmetric and time-independent metric for spacetime. For such a spacetime, the equation simplifies considerably since the only nonzero Christoffel symbols that contribute to the equation are $\Gamma_{00}^i$ and $\Gamma_{ij}^k$. The equation we get then reads,
\begin{equation}\label{ExplicitKGCST}
\frac{g^{00}}{c^2}\partial_t^2\psi-\frac{2im_ig^{00}}{\hbar}\partial_t\psi-\left(g^{00}\Gamma_{00}^k+g^{ij}\Gamma_{ij}^k\right)\partial_k\psi-\frac{m_i^2c^2}{\hbar^2}\left(g^{00}+1\right)\psi+g^{ij}\partial_{ij}\psi=0.
\end{equation}
Next, we need to pick up a metric for our spacetime. The obvious choice would be to use the Schwarzschild metric written in the Schwarzschild coordinates $(t,r,\theta,\phi)$. It turns out, however, that the Schwarzschild coordinates are not the best coordinates to use for our present purposes. The first reason is that with such a coordinate system, the wavefunction of the particle would have to be expanded in the spherical harmonics $Y^m_\ell(\theta,\phi)$ along the angular coordinates $\theta$ and $\phi$. Given that the wavefunction of the test particle does not wrap around the planet, but is rather limited in its extension to the size of the laboratory, a representation of the wavefunction in spherical harmonics is not recommended here. The second reason is that the spherical symmetry of the metric does not allow us to properly take into account the gravitational tidal forces in terms of local linear horizontal displacements from a given reference point. With such coordinates, one would have indeed to use angular displacements, which are not adequate for a table-top-size experiment.

Let us then use the Schwarzschild metric in the pseudo-Cartesian coordinates $(t,x,y,z)$ in which the metric reads (see, e.g., Ref.\,\cite{Catalog}),
\begin{align}\label{SchwarzschildDecarte}
{\rm d}s^2=&-\left(1-\frac{r_s}{r}\right)c^2{\rm d}t^2+\left[1+\frac{r_sx^2}{r^2(r-r_s)}\right]\!{\rm d}x^2\nonumber\\
&+\left[1+\frac{r_sy^2}{r^2(r-r_s)}\right]\!{\rm d}y^2+\left[1+\frac{r_sz^2}{r^2(r-r_s)}\right]\!{\rm d}z^2\nonumber\\
&+\frac{2r_s}{r^2(r-r_s)}\left(xy\,{\rm d}x{\rm d}y+xz\,{\rm d}x{\rm d}z+yz\,{\rm d}y{\rm d}z\right).
\end{align}
Here, $r_s=2GM/c^2$ is the Schwarzschild radius and $r=\sqrt{x^2+y^2+z^2}$ is the radial coordinate from the center of the Earth. It is easy now to extract from this metric the nonzero Christoffel symbols. Keeping only the leading-order terms, {\it i.e.}, those terms that contain at most one $c^2$ in the denominator, we are left with the following approximations for the inverse metric components $g^{\mu\nu}$ and for the Christoffel symbols $\Gamma_{\mu\nu}^\lambda$,
\begin{align}\label{MetricChristoffelComponents}
g^{00}&\approx-1-\frac{2GM}{c^2r}-\frac{4G^2M^2}{c^4r^2},\qquad
g^{ii}=1-\frac{2GM(x^i)^2}{c^2r^3},\nonumber\\[5pt]
g^{ij}&=-\frac{2GMx^ix^j}{c^2r^3}\;(i\neq j),\qquad
\Gamma_{00}^i\approx\frac{GMx^i}{c^2r^3},\nonumber\\[5pt]
\Gamma_{jj}^i&\approx\frac{GMx^i\left[2r^2-3(x^j)^2\right]}{c^2r^5},\qquad
\Gamma_{jk}^i\approx-\frac{3GMx^ix^jx^k}{c^2r^5}\;(j\neq k).
\end{align}
Note that we kept here the term with a factor of $1/c^4$ in the component $g^{00}$ because the latter also appears multiplied by $m_ic^2$ in one of the terms of Eq.\,(\ref{ExplicitKGCST}). Multiplying the whole equation (\ref{ExplicitKGCST}) by $\hbar^2/2m_i$, as well as by $g_{00}=1/g^{00}$, and then plugging the approximated terms (\ref{MetricChristoffelComponents}) into it, we get the following equation after keeping only those terms which are at least of the order of $1/c^2$:
\begin{equation}\label{KGPlugged}
\frac{\hbar^2}{2m_ic^2}\partial_t^2\psi-\frac{\hbar^2}{2m_i}\left[\delta^{ij}-\frac{2GM}{c^2r}\left(\delta^{ij}+\frac{x^ix^j}{r^2}\right)\right]\partial_{ij}\psi-\frac{\hbar^2}{m_i}\frac{GMx^i}{c^2r^3}\partial_i\psi-\frac{GMm_i}{r}\psi=i\hbar\partial_t\psi.
\end{equation}
We can now clearly distinguish from this expression the different contributions of the gravitational field. All the terms containing the $c^2$ in the denominator are relativistic corrections to the Newtonian gravitational potential $GM/r$. Among those relativistic terms, the ones that are proportional to $x^i$ and $x^ix^j$ are part of the relativistic tidal forces acting on the wavefunction of the particle. However, more tidal-force terms are still hiding inside the diagonal relativistic term $2GM\delta^{ij}/c^2r$, as well as inside the Newtonian term. To see this, one only needs to Taylor-expand the ratio $1/r$ to the second order in the Cartesian coordinates $(x,y,z)$. 

More important, however, is the presence of the cross-terms $x^ix^j$ that multiply the second derivatives $\partial_{ij}\psi$ of the wavefunction. Indeed, the presence of such terms prevents the wavefunction from being separable in the spatial coordinates. As a consequence, the wavefunction cannot anymore be split into the usual simple product $\psi(x,y,z)=\psi(x)\psi(y)\psi(z)$. The result is that the components of the wavefunction become spatially entangled\footnote{Spatial entanglement in the components of a wavefunction is not discussed often in literature. However, a very nice pedagogical exposition of the subject can be found in Ref.\,\cite{SchroederAjP}.}. We see then that even {\it classical} general relativity is able to induce a quantum entanglement (see, e.g., Refs.\,\cite{EPRinCST,BruschiNPj} for particle entanglement in curved-spacetime). The validity of this remark concerning the Newtonian gravitational potential will shortly become evident as well.         

In order to better estimate the orders of magnitude of each of the various terms in Eq.\,(\ref{KGPlugged}), and hence be able to better compare the relativistic terms to the nonrelativistic tidal forces, we need to consider concrete experimental values that could be involved in that equation. For definiteness, we consider the test particle to be a neutron. First, we have the Schwarzschild radius of the Earth that evaluates to $GM/c^2\approx4\,$mm. Therefore, using the radius of the Earth $R_E\approx6\times10^6\,$m yields $GM/(c^2R_E)\sim10^{-9}$. As a consequence, both the second and third terms inside the square brackets in Eq.\,(\ref{KGPlugged}) can safely be neglected. This is also true for the term that multiplies $\partial_i\psi$ in that equation. On the other hand, the very first term in that equation can be neglected because it is of the order of $\sim10^{17}$ times smaller than the kinetic term, even for a neutron. Therefore, the only source for possibly measurable tidal forces that remains inside this equation is the Newtonian potential. 

To investigate the effect of the latter in detail, let us perform the following change of variables $z\rightarrow R_E+z$. Therefore, the variable $z$ will henceforth represent in this section the upward vertical distance as measured from the surface of the Earth. As a consequence, we also have $r=\sqrt{(R_E+z)^2+x^2+y^2}$. As $x$, $y$ and $z$ are all of the size of the laboratory, and hence are all much smaller than $R_E$, we can Taylor-expand around the origin of the laboratory in each of the three coordinates. Keeping only terms up to the second order in $x/R_E$, $y/R_E$ and $z/R_E$, allows us to cast Eq.\,(\ref{KGPlugged}) into the following Schr\"odinger equation, 
\begin{equation}\label{TidalSchrodinger}
-\frac{\hbar^2}{2m_i}\left(\partial_{x}^2+\partial_{y}^2+\partial_{z}^2\right)\psi-\frac{GMm_g}{R_E}\left(1-\frac{z}{R_E}+\frac{z^2}{R_E^2}-\frac{x^2}{2R_E^2}-\frac{y^2}{2R_E^2}\right)\psi=i\hbar\partial_t\psi.
\end{equation}
Before we proceed, a couple of remarks are in order here. The first, is that we have introduced by hand the gravitational mass $m_g$ into Eq.\,(\ref{TidalSchrodinger}). We did this in view of discussing our results when they are thought of as emerging from a Newtonian gravity, which does allow such a distinction between the two masses. Whenever one wishes to focus instead only on what general relativity implies, one can always go back to the latter case by the replacement $m_g\rightarrow m_i$. The second remark is that Eq.\,(\ref{TidalSchrodinger}) is a differential equation which is separable in the coordinates $(x,y,z)$. Therefore, the solution to the equation might be found in the form of a product: $\psi(x,y,z)=\psi(x)\psi(y)\psi(z)$. However, had we kept at least the two third-order terms $zx^2/R_E^3$ and $zy^2/R_E^3$ in the equation, the latter would not have been separable anymore and the wavefunction would again exhibit a spatial entanglement in its components. This implies that even the classical Newtonian gravity can induce a quantum entanglement (see the recent Ref.\,\cite{Krisnanda} for work on a Newtonian gravity-induced particle entanglement).

Going back now to our equation (\ref{TidalSchrodinger}), we first notice that the constant term $-GMm_g/R_E$ would simply shift the whole energy spectrum of the particle by a constant. From what is left, we recognize the linear gravitational potential energy $-GMm_gz/R_E=-m_ggz$, where $g=GM/R_E^2$ is the gravitational acceleration on the surface of the Earth. The extra quadratic term $z^2/R_E$ could then be considered as a correction term to the linear potential. This suggests that we could first solve the unperturbed equation in $z$, which leads to a wavefunction given in terms of the square-integrable Airy function ${\rm Ai}(z)$, and then insert the latter into the time-independent perturbation theory to extract the corrected energy-spectrum for the freely-falling particle. A better and more insightful strategy, however, which will be useful to us in Sec.\,\ref{Sec:III} as well, consists instead in rearranging the terms in $z$ as follows:
\begin{equation}\label{RearrangedSchrodinger}
-\frac{\hbar^2}{2m_i}\left(\partial_{x}^2+\partial_{y}^2+\partial_{\xi}^2\right)\psi+\frac{m_g}{2}\omega^2\left(x^2+y^2-2\xi^2\right)\psi-\kappa\psi=i\hbar\partial_t\psi.
\end{equation}
In writing this equation, we introduced the new variable $\xi=\frac{1}{2}R_E-z$ and gathered the constant terms on the left-hand side into the single constant $\kappa=\frac{3}{4}GMm_g/R_E$. We also introduced the constant angular frequency $\omega=\sqrt{g/R_E}$. This corresponds to an oscillation period of about one hour on the surface of the Earth. Using this angular frequency, we then also have $\kappa=\frac{3}{4}m_g\omega^2R_E^2$. The form of the gravitational potential in the Schr\"odinger equation (\ref{RearrangedSchrodinger}) is that of a simple harmonic oscillator along the horizontal direction, spanned by the coordinates $(x,y)$. However, along the vertical direction, the gravitational potential is that of the well-known inverted harmonic ---\! sometimes also called repulsive\! --- oscillator \cite{Barton,Yuce,Munoz}, centered at $\xi=0$. Of course, this center of the inverted harmonic oscillator is only a fictitious position, for in our approximations we have $z\ll R_E$. 

In order to solve the inverted harmonic oscillator equation governing the dynamics of the component $\psi(\xi,t)$ along the vertical direction, we have actually two options. The first option consists in leaving the system to evolve in a time-independent way. In such a case, the wavefunction will also be time-independent and will be dealt with in Sec.\,\ref{Sec:II.a}. The second option consists in imposing on the system time-dependent boundary conditions. The wavefunction in the latter case will obviously be time-dependent as well, and will be dealt with in Sec.\,\ref{Sec:II.b}.
\subsection{Using a time-independent boundary condition}\label{Sec:II.a}
In case a single time-independent boundary condition
is applied in the vertical direction, the system will be time-independent as well. We can thus cast the constant $\kappa$ in the Schr\"odinger equation (\ref{RearrangedSchrodinger}) into the right-hand side and absorb it into the particle's constant energy $\mathcal{E}$ on the right-hand side by redefining the ground-level energy. The separability of the equation implies then that the total energy $E$ of the particle is made of three pieces as well: $E=E_{n_x}+E_{n_y}+\mathcal{E}$. The usual textbook treatment of the time-independent harmonic oscillator in one dimension (see, e.g., Ref.\,\cite{Greiner}) easily yields the two pieces $E_{n_x}$ and $E_{n_y}$ arising from the horizontal harmonic motion as follows:
\begin{equation}\label{xyHarmonicEnergy}
E_{n_x}+E_{n_y}=\hbar\omega\sqrt{\frac{m_g}{m_i}}\left(n_x+n_y+1\right),
\end{equation}
where, $n_x$ and $n_y$ are two non-negative integers. The ratio $m_g/m_i$ emerges here due to the different masses attached to the kinetic and potential terms of the harmonic oscillators. This result is very interesting because, just as in Newtonian mechanics, we remarkably obtained the ratio $m_g/m_i$ in an isolated form even within quantum mechanics. The precision in measuring this ratio is thus directly linked to the precision that could be achieved in measuring the frequency $\omega$ of the oscillator. When one is assuming the framework of general relativity, one should set $m_g=m_i$. In that case, Eq.\,(\ref{xyHarmonicEnergy}) implies a mass-independent energy spectrum in the $xy$-plane. In order to find the energy $\mathcal{E}$, we should solve the $\xi$-equation.

The resulting differential equation in $\xi$ along the vertical direction has the form of Weber's equation \cite{FormulasBook1}. In order to simplify our mathematical manipulations, it is more convenient to introduce a scaling factor $\ell$, given by $\ell=[\hbar^2R_E/(2m_im_gg)]^{\frac{1}{4}}$. With the help of this scaling factor, we can define the dimensionless position parameter $\zeta=\xi/\ell$, as well as the dimensionless energy parameter $\eta=\mathcal{E}R_E/(2m_gg\ell^2)$. Substituting these redefinitions into Eq.\,(\ref{RearrangedSchrodinger}), the equation obeyed by $\psi(\zeta)$ reads
\begin{equation}\label{Schrodinger}
-\frac{1}{2}\left(\frac{{\rm d}^2}{{\rm d}\zeta^2}+\zeta^2\right)\psi=\eta\,\psi.
\end{equation}
The solution of such an equation provides the well-known eigenstates \cite{Barton} of the time-independent inverted harmonic oscillator in terms of the parabolic cylinder function. Those eigenstates constitute a complete orthonormal basis in the Hilbert space $\mathcal{L}^2(R)$ of square-integrable functions, but in the Dirac sense (see, e.g., Ref.\,\cite{EngineeringBook}). The eigenstates thus normalized read \cite{EngineeringBook},
\begin{equation}\label{CylinderSolution}
\psi(\zeta)=\frac{e^{i\frac{\pi}{8}+\frac{\pi}{4}\eta}}{2^{\frac{3}{4}}\pi}\,\Gamma\left(\frac{1}{2}-i\eta\right)\mathcal{D}_{i\eta-\frac{1}{2}}\left(\pm e^{i\frac{3\pi}{4}}\sqrt{2}\,\zeta\right).
\end{equation}
Here, $\Gamma(a)$ is the gamma function for any complex constant $a$, and $\mathcal{D}_{ib-\frac{1}{2}}(\sigma)$ is the Whittaker form of the parabolic cylinder function for any complex variable $\sigma$ and for any real constant $b$ \cite{FormulasBook2}. For us, the important fact about this solution is that the parabolic cylinder function exhibits the remarkable feature that when its constant parameter $b$ is large, it admits an expansion in terms of Airy functions \cite{FormulasBook2,Olver} for all positive values of the variable $\zeta$, including the case for which $\zeta\gg1$. 

It turns out that this is exactly our case. Indeed, recall that the definition of our dimensionless variable is $\zeta=\xi/\ell=(\frac{1}{2}R_E-z)/\ell$. Also, for a neutron particle, for example, we find that the scaling factor $\ell$ is of the order of $\sim6\,$mm. Therefore, close to the ground level at $z=0$, we have $\zeta\gg1$. On the other hand, from the definition of our dimensionless energy parameter $\eta$, we also have $|\eta|=|\mathcal{E}|R_E/(2m_gg\ell^2)\gg1$. This stems from the fact that, according to Eq.\,(\ref{Schrodinger}), the parameter $\eta$ is made of the kinetic energy of the particle and the term $-\frac{1}{2}\zeta_0^2$, where $\zeta_0=\frac{1}{2}R_E/\ell$ is obtained by setting $z=0$ in $\zeta$. In other words, $\mathcal{E}$ in Eq.\,(\ref{Schrodinger}) is now made of the gravitational potential energy $-\frac{1}{4}m_ggR_E$ plus a comparatively small positive kinetic energy $\epsilon$. Furthermore, the condition $\zeta/\sqrt{2|\eta|}>1$ that we are in brings us to a very specific expansion in terms of the Airy function ${\rm Ai}(z)$ \cite{FormulasBook2,Olver}. Using such an expansion, we find the following expression of the wavefunction for all $\zeta$, {\it i.e.}, for all $z$, inside the laboratory stuck to the ground, 
\begin{equation}\label{CylinderAsymptotic}
\psi(\zeta)\sim2^{\frac{3}{4}}\sqrt{\pi}\,e^{\frac{\pi}{2}|\eta|}\left(\frac{u}{\zeta^2-2|\eta|}\right)^{\frac{1}{4}}{\rm Ai}(-u).
\end{equation}
The variable $u$ in this expression is given in terms of our $\zeta$ and $|\eta|$ by \cite{FormulasBook2},
\begin{equation}\label{uVariable}
u=\left(\frac{3}{2}|\eta|\right)^{\frac{2}{3}}\left(\frac{\zeta}{\sqrt{2|\eta|}}\sqrt{\frac{\zeta^2}{2|\eta|}-1}-\arccosh\frac{\zeta}{\sqrt{2|\eta|}}\right)^{\frac{2}{3}}.
\end{equation}
From expression (\ref{CylinderAsymptotic}), we see that the only way for the wavefunction to satisfy the boundary condition $\psi(z=0)=0$, is to have the Airy function vanish at $z=0$, for which the variable $\zeta$ reduces to $\zeta_0$. On the other hand, we know that, to a good approximation, the zeros of the Airy function ${\rm Ai}(-u)$ are given by $u_0\approx[\frac{3\pi}{2}(n-\frac{1}{4})]^{\frac{2}{3}}$, where $n$ is a non-negative integer  \cite{FormulasBook2}. Substituting expression (\ref{uVariable}) of $u$ into these zeros of the Airy function leads to a transcendental equation involving $|\eta|$, $\zeta_0$ and the integer $n$. Thus, an exact analytic expression for $|\eta|$, {\it i.e.} for the kinetic energy $\epsilon$ of the particle, in terms of $\zeta_0$ and $n$ cannot be obtained. What we can do instead, is write $|\eta|=\frac{1}{2}\zeta_0^2-\epsilon R_E/(2m_gg\ell^2)$, and expand in $\epsilon$ the right-hand side of Eq.\,(\ref{uVariable}) up to the first nonvanishing order in $\epsilon$, which is $\epsilon^{\frac{3}{2}}$. The zeros $u_0$ of the Airy function then lead to the following constraint on $\epsilon$ in order for the boundary condition $\psi(z=0)=0$ to be satisfied:
\begin{equation}\label{EtaDiscretization}
    \epsilon\approx\frac{2m_gg\ell^2}{R_E}\left[\frac{3\pi\zeta_0}{2\sqrt{2}}\left(n-\frac{1}{4}\right)\right]^{\frac{2}{3}}.
\end{equation}
By substituting our definitions of the parameters $\ell$ and $\zeta_0$ into this result, we recover exactly expression (\ref{AiryQuantization}) obtained for the particle inside the linear gravitational field of the Earth. 

Let us now pause here to ponder on the way we arrived at this result and how it compares to the result (\ref{AiryQuantization}) obtained with the linear-field approximation of the gravitational field of the Earth. First, we should note that the energy spectrum of the inverted harmonic oscillator in the $z$-direction gave rise to the same spectrum as the one implied by the linear potential thanks to the fact that, far from its fictitious center $z_0=\frac{1}{2}R_E$, the shape of the inverted harmonic oscillator potential tends to the linear potential. Nevertheless, our approach, which consists in finding the exact eigenstates of the Hamiltonian in Eq.\,(\ref{TidalSchrodinger}) instead of considering the term proportional to $z^2$ as a mere perturbing term, has been very instructive. Indeed, our approach showed that although the ratio $m_g/m_i$ appears in the wavefunction (\ref{CylinderSolution}) inside the parameter $\eta$, such a ratio does not appear in an isolated form. The presence of $\ell$ inside $\zeta$ is what contaminates such a ratio. This is in contrast to the gravitational harmonic oscillators along the horizontal directions $x$ and $y$ for which the ratio $m_g/m_i$ remained pure. Moreover, the probability for finding the particle at a certain height $z$ is given by $|\psi(z)|^2$ which, according to Eq.\,(\ref{CylinderAsymptotic}), is oscillating thanks to the Airy function like in the linear case, but with an amplitude which is here contained inside an envelop that is increasing with the height $z$ like $(\frac{1}{2}R_E-z)^{-\frac{1}{3}}$. Furthermore, even this envelope depends on the ratio $m_g/m_i$ in a non-isolated form.

The other very important fact we would like to emphasize now is that even by setting $m_g=m_i$, as required by general relativity, the energy spectrum (\ref{EtaDiscretization}) does not become mass-independent as is the case for the harmonic oscillators along the horizontal directions. In other words, even though both kinds of harmonic oscillators are gravitationally induced, one of them becomes mass-independent while the other remains mass-dependent when both concepts of mass coincide. We shall come back to this observation in the conclusion section. 
\subsection{Using time-dependent boundary conditions}\label{Sec:II.b}
As we did in the previous subsection, in order to solve Eq.\,(\ref{RearrangedSchrodinger}) we only need to split the wavefunction into the product $\psi(x,y,\xi,t)=\psi(z)\psi(y)\psi(\xi,t)$, and solve separately the three resulting independent equations. The reason for the time dependence of the third component, and for not having discarded the constant $\kappa$ as is customary when dealing with the time-independent Schr\"odinger equation, is due to the time dependence of the boundary conditions that we are going to impose here on the wavefunction along the vertical direction. This approach admits an exact analytic treatment thanks to the technique introduced in Ref.\,\cite{Yuce}. We shall expose here in detail the steps behind such a technique by adapting it to our case, which is slightly different from the one exposed in Ref.\,\cite{Yuce}. 

We start by making the following successive redefinitions, first a redefinition of the wavefunction $\psi(\xi,t)$, and then followed by a redefinition of the variable $\xi$:
\begin{align}
\psi(\xi,t)&=\exp\left(\frac{i\omega\sqrt{m_im_g}}{\sqrt{2}\hbar} \xi^2+\frac{i\kappa}{\hbar}t-\sqrt{\frac{m_g}{2m_i}}\omega t\right)\Phi(\xi,t),\label{Redefintion1}\\
\xi&=u\,e^{\sqrt{\frac{2m_g}{m_i}}\omega t}.\label{Redefintion2}    
\end{align}
Inserting the redefinition (\ref{Redefintion1}) into the Schr\"odinger equation for $\psi(\xi,t)$ as given by Eq.\,(\ref{RearrangedSchrodinger}), the differential equation one obtains for $\Phi(\xi,t)$ takes the following form:
\begin{equation}\label{SchrodingerRedefined}
-\frac{\hbar^2}{2m_i}\frac{\partial^2\Phi}{\partial\xi^2}-i\hbar\omega\sqrt{\frac{2m_g}{m_i}}\,\xi\frac{\partial\Phi}{\partial \xi}=i\hbar\frac{\partial\Phi}{\partial t}.
\end{equation}
Finally, with the redefinition (\ref{Redefintion2}) of the variable $\xi$, we have $\partial_\xi=e^{-\sqrt{2m_g/m_i}\,\omega t}\,\partial_u$ and we have to turn the operator $\partial_t$ in Eq.\,(\ref{SchrodingerRedefined}) into the operator $\partial_t- \sqrt{2m_g/m_i}\,\omega u\,\partial_u$. 
The differential equation for the new wavefunction $\Phi(u,t)$, which thus becomes a function of the variables $u$ and $t$, takes then the following form:
\begin{equation}\label{PHIEquation}
-\frac{\hbar^2}{2m_i}\frac{\partial^2\Phi}{\partial u^2}=i\hbar e^{\sqrt{\frac{8m_g}{m_i}}\,\omega t}\,\frac{\partial\Phi}{\partial t}.
\end{equation}
This equation is easily solvable as it is separable in the variables $u$ and $t$. Two sets of linearly independent solutions are possible. Introducing a positive constant $\epsilon$ with the dimensions of energy, and setting $\varepsilon=\sqrt{2m_i\epsilon/\hbar^2}$, the two possible solutions read
\begin{equation}
\Phi_1(u,t)\!=\!\exp\left({\frac{i\epsilon}{\hbar\omega}\!\sqrt{\frac{m_i}{8m_g}}e^{-\sqrt{\frac{8m_g}{m_i}}\,\omega t}}\right)\left[C_1\sin\left(\varepsilon u\right)\!+\!C_2\cos\left(\varepsilon u\right)\right],\label{4Solutions1}
\end{equation}
\begin{equation}
\Phi_2(u,t)=\exp\left({\frac{i\epsilon}{\hbar\omega}\sqrt{\frac{m_i}{8m_g}}e^{-\sqrt{\frac{8m_g}{m_i}}\,\omega t}}\right)\left(C_3e^{-\varepsilon u}+C_4e^{+\varepsilon u}\right).\label{4Solutions2}    
\end{equation}
Note the important appearance, just as in Eq.\,(\ref{xyHarmonicEnergy}), of the ratio $m_g/m_i$ in an isolated form in both solutions. The only solution which we can obviously discard out of hand, as being unphysical, is the second term which is proportional to $e^{+\varepsilon u}$ in the combination (\ref{4Solutions2}). On the other hand, the first term that multiplies $C_3$ in that combination (\ref{4Solutions2}) is normalizable. The reason why we cannot take that one as a physical solution either is that this technique requires, in addition, to impose the vanishing of the wavefunction at two distinct boundaries. It is obvious that the latter requirement cannot be imposed on the first term multiplying $C_3$ in Eq.\,(\ref{4Solutions2}), which does not vanish for any finite value of $u$ unless we set $C_3=0$. We have then no other choice, but to consider only the solution given by Eq.\,(\ref{4Solutions1}). 

Both terms in the combination (\ref{4Solutions1}) do not yield normalizable wavefunctions all over space in the variable $\xi$. However, as pointed out in Ref.\,\cite{Yuce}, such a solution can be normalized in the variable $u$ when the wavefunction $\Phi_1(u,t)$ is confined between the origin at $u=0$, {\it i.e.}, at $\xi=0$, and some other position $u=h_0$, for some constant $h_0$. This means that the wavefunction $\Phi_1(u,t)$ has to vanish at these two boundaries. In our case, the obvious first boundary would be at $z=0$. Unfortunately, $z=0$ corresponds to $\xi=\frac{1}{2}R_E$ for which the combination (\ref{4Solutions1}) cannot vanish either, unless $C_1=C_2=0$. Therefore, in order to adapt that strategy to our case, we choose the distinct boundaries to be at $u=h_0$ and $u=h_1$, both different from the origin $u=0$. In terms of the variable $z$, these time-dependent positions are realized at $z_0=\frac{1}{2}R_E-h_0\exp(\sqrt{\frac{2m_g}{m_i}}\omega t)$ and at $z_1=\frac{1}{2}R_E-h_1\exp(\sqrt{\frac{2m_g}{m_i}}\omega t)$, respectively. Imposing $\Phi_1(h_0,t)=0$ on the combination (\ref{4Solutions1}), we find that $C_2=-C_1\tan(\varepsilon h_0)$. Next, substituting $C_2$ in terms of $C_1$ into the condition $\Phi_1(h_1,t)=0$, we arrive at the following single condition on $\varepsilon$:
\begin{equation}\label{EpsilonQuantization}
\sin\left[\varepsilon(h_1-h_0)\right]=0.
\end{equation}
This condition is satisfied if and only if $\varepsilon=n\pi/(h_1-h_0)$, for any integer $n$.

The last step now is to normalize the wavefunction $\psi(\xi,t)$ as given by Eq.\,(\ref{4Solutions1}), by demanding that $\int|\psi(\xi,t)|^2{\rm d}\xi$ evaluates to unity over the interval of the confinement \cite{Yuce}. First, inserting the combination (\ref{4Solutions1}) into expression (\ref{Redefintion1}), and taking into account Eq.\,(\ref{Redefintion2}), as well as the relation we just found between $C_1$ and $C_2$, we arrive at the following final expression for the wavefunction $\psi(\xi,t)$:
\begin{align}\label{PsiSolution}
\psi(\xi,t)&=
C_1\exp\left(\frac{i\omega\sqrt{m_im_g}}{\sqrt{2}\hbar} \xi^2\!+\!\frac{i\kappa}{\hbar}t\!-\!\sqrt{\frac{m_g}{2m_i}}\omega\,t\!+\!\frac{i\epsilon}{\hbar\omega}\sqrt{\frac{m_i}{8m_g}}e^{-\sqrt{\frac{8m_g}{m_i}}\,\omega t}\right)\nonumber\\
&\qquad\times\left[\sin\left(\varepsilon e^{-\sqrt{\frac{2m_g}{m_i}}\omega t}\xi\right)-\tan\left(\varepsilon h_0\right)\cos\left(\varepsilon e^{-\sqrt{\frac{2m_g}{m_i}}\omega t}\xi\right)\right].    
\end{align}
The normalization of this wavefunction in the interval $u\in[h_0,h_1]$ corresponds to the normalization in the variable $\xi$ within the interval $\xi\in[h_0\exp(\sqrt{\frac{2m_g}{m_i}}\omega t), h_1\exp(\sqrt{\frac{2m_g}{m_i}}\omega t)]$. Evaluating the integral leads to the following result after rearranging the terms and using the condition (\ref{EpsilonQuantization}):
\begin{equation}\label{Normalization}
\int_{h_0\exp(\sqrt{\frac{2m_g}{m_i}}\,\omega t)}^{h_1\exp(\sqrt{\frac{2m_g}{m_i}}\,\omega t)}|\psi(\xi,t)|^2{\rm d}\xi=|C_1|^2\left(\frac{h_1-h_0}{2\cos^2(\varepsilon h_0)}-\left[\frac{\cos(\varepsilon h_1)}{4\varepsilon\cos(\varepsilon h_0)}-\frac{\tan^2(\varepsilon h_0)}{2\varepsilon}\right]\sin\left[\varepsilon(h_1+h_0)\right]\right).
\end{equation}
In order to simplify this expression even further, we may assume that the distances $h_0$ and $h_1$ are chosen such that $h_0+h_1=s\,(h_1-h_0)$, with a positive integer number $s$. In fact, in such a case, $\sin[\varepsilon(h_1+h_0)]$ vanishes, and only the first term inside the parentheses in Eq.\,(\ref{Normalization}) would remain. Therefore, by choosing the relative heights such that $h_1=\frac{s+1}{s-1}h_0$, for any integer $s>1$, we end up with the following normalization constant for the wavefunction:
\begin{equation}\label{FinalNormalization}
C_1=\sqrt{\frac{s-1}{h_0}}\cos(\varepsilon h_0).
\end{equation}
We would like to emphasize here that Eq.\,(\ref{Normalization}) is valid for the general case, and can be used to extract the normalization constant in that general case as well. However, to achieve our present purposes, it is sufficient to consider only the simplifying special case (\ref{FinalNormalization}) which does not compromise in any way the generality of the result we obtain below.

In fact, after plugging this value of the normalization constant into expression (\ref{PsiSolution}) of the wavefunction, as well as the condition (\ref{EpsilonQuantization}), we compute the energy spectrum of the particle, up to an unimportant overall constant, using the Hamiltonian $\mathcal{H}$ of the inverted harmonic oscillator as follows,
\begin{align}\label{Hamiltonian}
E_{n_\xi}&=\int_{h_0\exp\left(\sqrt{\frac{2m_g}{m_i}}\,\omega t\right)}^{h_1\exp\left(\sqrt{\frac{2m_g}{m_i}}\,\omega t\right)}\psi^*(\xi,t)\,\mathcal{H}\,\psi(\xi,t)\,{\rm d}\xi\nonumber\\[3pt]
&=\frac{n^2\pi^2\hbar^2|C_1|^2}{4(h_1-h_0)m_i\cos^2(\varepsilon h_0)}e^{-\sqrt{\frac{8m_g}{m_i}}\omega t}\nonumber\\[3pt]
&=\frac{n^2(s-1)^2\pi^2\hbar^2}{8m_ih_0^2}e^{-\sqrt{\frac{8m_g}{m_i}}\omega t}.    
\end{align}
The second line in this result is what one obtains for the general case when there is no special link between the experimental heights $h_1$ and $h_0$ chosen for the boundaries. The last line is what one obtains for the special case $h_1=\frac{s+1}{s-1}h_0$, for any integer $s>1$, for which the formal expression simplifies greatly. In both cases, however, the general pattern displayed by the particle's energy is that the latter is quantized and decays exponentially with time, with a decay rate given by $\sqrt{\frac{8m_g}{m_i}}\omega$. The remarkable dependence of this energy decay on the ratio $m_g/m_i$, instead of just a dependence on either $m_i$ or $m_g$, is due to the fact that we forced the system to react simultaneously to a stimulation caused by the gravitational field and another stimulation that is kinematical in nature. As a consequence, the system has to react with its gravitational mass to the first and has to react with its inertial mass to the second. The presence of the gravitational stimulus gives rise to the ratio $m_g/m_i$, whereas the inertial stimulus gives rise to the isolated mass $m_i$ in the denominator. For this reason, even if one sets $m_g=m_i$ to conform to the framework of general relativity, the mass dependence in the energy spectrum (\ref{Hamiltonian}) would still remain. 

Given the smallness of $\omega$, it is clear that the decay rate of the energy is very small too. Given that the motion of the two time-varying boundaries to be imposed on the particle needs to be exponentially accelerating at a rate proportional to $\omega$, an eventual experimental realization of the setup requires achieving long-lived quantum coherence.

\section{Tidal forces in the inertial frame}\label{Sec:III}
Our investigation in the previous section allowed us to extract the effect of tidal forces on the wavefunction in two different settings, both as seen from a noninertial frame. To achieve that, we made use of the Schwarzschild metric (\ref{SchwarzschildDecarte}) as perceived by the noninertial observer stuck to the ground. To describe the local physics inside a freely-falling laboratory, we need to use, instead, the Schwarzschild metric written in Fermi normal coordinates $(t,x,y,z)$. In such coordinates, the metric takes the following form along a geodesic path parallel to the $z$-axis \cite{Manasse}:  \begin{align}\label{SchwarzschildFermi}
{\rm d}s^2=&-\left[1+\frac{r_s}{2r^3(\tau)}\left(x^2+y^2-2z^2\right)\right]c^2{\rm d}t^2\nonumber\\
&+\left[1\!+\!\frac{r_s}{6r^3(\tau)}\left(z^2-2y^2\right)\right]\!{\rm d}x^2+\left[1\!+\!\frac{r_s}{6r^3(\tau)}\left(z^2-2x^2\right)\right]\!{\rm d}y^2\nonumber\\
&+\left[1\!+\!\frac{r_s}{6r^3(\tau)}\left(x^2+y^2\right)\right]\!{\rm d}z^2\nonumber\\
&-\frac{r_s}{3r^3(\tau)}\left(xz\,{\rm d}x{\rm d}z+yz\,{\rm d}y{\rm d}z-2xy\,{\rm d}x{\rm d}y\right).
\end{align}
This metric is time-dependent through the dependence of the radial coordinate $r(\tau)$ on the proper time $\tau$ as measured by an observer on board the freely-falling laboratory. For such a spacetime, the components $g_{0i}$, and hence also the components $g^{0i}$, still vanish and the only nonzero Christoffel symbols that would be required to write down the Klein-Gordon equation are $\Gamma_{00}^0$, $\Gamma_{00}^i$, $\Gamma_{ij}^0$ and $\Gamma_{ij}^k$. Therefore, Eq.\,(\ref{KGCST}) takes the following explicit form,
\begin{multline}\label{ExplicitKGCST2}
\frac{g^{00}\hbar^2}{2m_ic^2}\partial_t^2\psi-i\hbar\left(g^{00}+\frac{\hbar}{2m_ic}g^{ij}\Gamma_{ij}^0+\frac{i\hbar}{2m_ic}g^{00}\Gamma_{00}^0\right)\partial_t\psi\\-\frac{\hbar^2}{2m_i}\left(g^{00}\Gamma_{00}^k+g^{ij}\Gamma_{ij}^k\right)\partial_k\psi+\frac{\hbar^2}{2m_i}g^{ij}\partial_{ij}\psi\\-\frac{m_ic^2}{2}\left(g^{00}+1-\frac{i\hbar}{m_ic}g^{00}\Gamma_{00}^0-\frac{i\hbar}{m_ic}g^{ij}\Gamma_{ij}^0\right)\psi=0.
\end{multline}

Having already checked the orders of magnitude of the majority of the relativistic terms in this equation in the previous section, and found that they are at least $1/c^2$ smaller than the tidal-force terms, we only need to focus here on the new terms containing $\Gamma_{00}^0$ and $\Gamma_{ij}^0$. The forms of these Christoffel symbols are readily evaluated to be $\Gamma_{00}^0=\frac{1}{2c}g^{00}\partial_tg_{00}$ and $\Gamma_{ij}^0=-\frac{1}{2c}g^{00}\partial_tg_{ij}$. Therefore, all the new terms in Eq.\,(\ref{ExplicitKGCST2}) are proportional to $1/c^3$ except the last term inside the parentheses in the last line, which has the form $\frac{i\hbar}{4}g^{ij}g^{00}\partial_tg_{ij}$. This term is of the order of $GM(x^i)^2/c^2r^4$ and can therefore also be ignored together with the terms we were able to drop from the previous section. Substituting the components of the inverse metric extracted from Eq.\,(\ref{SchwarzschildFermi}), and keeping only the leading terms, the differential equation (\ref{ExplicitKGCST2}) takes the following form:
\begin{equation}\label{FermiNormalSchrodinger}
-\frac{\hbar^2}{2m_i}\left(\partial_{x}^2+\partial_{y}^2+\partial_{z}^2\right)\psi+\frac{GMm_i}{2r^3(\tau)}\left(x^2+y^2-2z^2\right)\psi=i\hbar\partial_\tau\psi.
\end{equation}
We have introduced here the proper-time derivative $\partial_\tau\psi$ by noting that $\partial_t\psi=\frac{{\rm d}\tau}{{\rm d} t}\partial_\tau\psi$ and that $\frac{\partial\tau}{\partial t}\approx1-r_s/r\approx1$. Note, also, that we have not introduced
here the gravitational mass $m_g$ by hand as we did previously. The reason is that the time-dependence of the system we are going to obtain yields an expression that is already very involved even when identifying the two masses. Moreover, as we shall discuss below, in contrast to what we did so far, the appearance of the ratio $m_g/m_i$ in the final results in this section would not be altered much by ignoring the possibility of having $m_g\neq m_i$. Indeed, as we shall see, our results bring a new perspective by specifically setting $m_i=m_g$ and examining the outcome. 

Once more, we recognize in Eq.\,(\ref{FermiNormalSchrodinger}) the simple harmonic potentials along the horizontal $xy$-plane and the inverted harmonic oscillator along the vertical $z$-axis. In other words, we have a generic distinction between the vertical direction and the horizontal directions. The tidal forces along the former are due to the inverse-square law of gravity, whereas tidal forces along the latter are due to the geometric shape of the mass source. The difference with the previous section, however, is that here both potentials are time-dependent. The other difference is that the three gravitational potentials have now the same center along the three spatial directions. In addition, the wavefunctions are again separable into three independent pieces along the three different spatial directions. As such, and since finding the energy spectrum and the eigenstates of the inverted harmonic oscillator becomes much more involved when attempting to solve the Schr\"odinger equation exactly the way we did it in the previous section, we only need to focus on one of the two time-dependent simple harmonic oscillators in the $xy$-plane. 

To be able to solve Eq.\,(\ref{FermiNormalSchrodinger}), we need first to examine the explicit expression of $r^3(\tau)$, which is given by solving the radial geodesic equation in the Schwarzschild spacetime\footnote{See, e.g., Ref.\,\cite{Lambourne} where the derivations leading to the radial geodesic equation in the Schwarzschild metric needed here is worked out in much detail.}. Within our approximation of nonrelativistic motion, we have $E\approx m_ic^2$, in which case the solution of the radial geodesic equation reduces to the Newtonian result,
\begin{equation}\label{rCube}
r^3(\tau)=\frac{9}{2}GM\left(\sqrt{\frac{2r_0^3}{9GM}}-\tau\right)^2.
\end{equation}
We have chosen the constant of integration such that $r=r_0$ at $\tau=0$. Note, in addition, that this simple expression (\ref{rCube}) is obtained by assuming, either the reference frame is released without initial speed infinitely far away from the center of the mass source, or else, the non-zero initial speed $\sqrt{2GM/r_0}$ of the reference frame would have to be considered instead. This is the price to pay to avoid having our subsequent formulas in this section become unwieldy. 

We need now to estimate the time duration required for the experiment in order for the time-dependent term in this expression to make a non-negligible influence on the time-evolution of the energy of the harmonic oscillator. If we take the Earth to be the mass source, and take $r_0$ to be of the order of Earth's radius $R_E$, we find that $\sqrt{2r_0^3/9GM}\approx380\,$seconds. Therefore, for the time-dependence to make a significant contribution to the shaping of the energy of the harmonic oscillator, the duration of the free fall needs to be at least around $38$ seconds. If, on the other hand, we consider instead initial distances $r_0$ which are much larger than the Earth's radius in order to avoid the initial unrealistic large speeds, than the duration of the free fall would, of course, be much larger than these $38\,$seconds. We shall come back to this issue in the conclusion section. We are now going to find the energy spectrum of the particle from Eq.\,(\ref{FermiNormalSchrodinger}).

First, we introduce here the decreasing reduced time $\bar{t}=\bar{t}_0-\tau$ for convenience, where we have set $\bar{t}_0=(2r_0^3/9GM)^{\frac{1}{2}}$. Then, we separate the total wavefunction into $\psi(x,y,z,\bar{t})=\chi(x,\bar{t})\upsilon(y,\bar{t})\zeta(z,\bar{t})$ and consider here only the $x$-component $\chi(x,\bar{t})$. The one-dimensional Schr\"odinger equation satisfied by this component after substituting Eq.\,(\ref{rCube}) into (\ref{FermiNormalSchrodinger}), is given by,
\begin{equation}\label{ChiSchrodinger}
-\frac{\hbar^2}{2m_i}\partial_{x}^2\chi+\frac{m_i}{9\bar{t}^2}x^2\chi=-i\hbar\partial_{\bar{t}}\chi.
\end{equation}
Note also that the proper time $\tau$ increases from $0$, for which $\bar{t}=\bar{t}_0$, and tends to $\bar{t}_0$, for which both $\bar{t}$ and $r(\tau)$ vanish. The time scale $\bar{t}_0$ should then be taken here as an asymptotic value for $\tau$ because the freely-falling laboratory frame does not go to $r=0$. The minimum distance from the center of the Earth that the laboratory can reach is the radius of the Earth. Now, this equation (\ref{ChiSchrodinger}) has the well-known form of a general time-dependent frequency harmonic oscillator, the time-dependent energy spectrum of which is given in Refs.\,\cite{Lewis1967,Lewis1969}. Using the results provided in the latter references, Eq.\,(\ref{ChiSchrodinger}) yields the following expectation value of the Hamiltonian along the $x$-direction:
\begin{equation}\label{ExpectationValue}
\bra{n}{H_x}\ket{n}=\frac{\hbar\left(n+\frac{1}{2}\right)}{2m_i}\left[m_i^2{\dot\rho}^2(\bar{t})+\frac{2m_i^2}{9\bar{t}^2}\rho^2(\bar{t})+\frac{1}{\rho^2(\bar{t})}\right].
\end{equation}
Here, an overdot denotes a derivative with respect to $\bar{t}$, and  $\ket{n}$ are normalized eigenstates with $n$ a non-negative integer \cite{Lewis1967,Lewis1969}. The auxiliary function of time $\rho(\bar{t})$ satisfies a second-order and nonlinear differential equation that reads,
\begin{equation}\label{RhoEquation}
m_i^2\ddot{\rho}+\frac{2m_i^2}{9\bar{t}^2}\rho=\frac{1}{\rho^3}.
\end{equation}
By solving this differential equation we can immediately find the evolution in time of the energy spectrum as given by the expectation values (\ref{ExpectationValue}) for each eigenstate $n$. In addition, however, the Hamiltonian $H_x$ has also off-diagonal matrix elements. As a consequence, the off-diagonal matrix elements $\bra{n'}H\ket{n}$ are nonzero whenever $n'=n\pm2$ \cite{Lewis1969}. Being interested here in the diagonal matrix elements, we restrict our focus on the latter.

To solve Eq.\,(\ref{RhoEquation}), which is the Ermakov equation, we follow the method outlined in Ref.\,\cite{FormulasBook1} for solving such an equation. The derivation of the solution $\rho(\bar{t})$ is quite long, so the detailed steps are given in Appendix \ref{A}. Plugging the result (\ref{Rho2FinalFinal}) for $\rho(\bar{t})$ and $\dot{\rho}(\bar{t})$ into Eq.\,(\ref{ExpectationValue}), we arrive at the following evolution in time of the expectation value of the Hamiltonian:
\begin{equation}\label{ExpectationValueFinal}
\bra{n}{H_x}\ket{n}=\frac{\hbar}{\sqrt{2}\bar{t}_0}\left(n+\frac{1}{2}\right)\frac{9\left(\frac{\bar{t}_0}{\bar{t}}\right)^{2}-35\left(\frac{\bar{t}_0}{\bar{t}}\right)^{\frac{5}{3}}+\frac{643}{12}\left(\frac{\bar{t}_0}{\bar{t}}\right)^{\frac{4}{3}}-\frac{75}{2}\left(\frac{\bar{t}_0}{\bar{t}}\right)+\frac{81}{8}\left(\frac{\bar{t}_0}{\bar{t}}\right)^{\frac{2}{3}}}{3\left(\frac{\bar{t}_0}{\bar{t}}\right)^{\frac{2}{3}}-5\left(\frac{\bar{t}_0}{\bar{t}}\right)^{\frac{1}{3}}+\frac{9}{4}}.
\end{equation}
We clearly see from this time evolution of the expectation value the effect of the tidal forces on the free fall. As the particle comes closer to the surface of the Earth, the tidal forces become stronger. The induced harmonic oscillations become, as a consequence, stronger as well. Indeed, as the proper time $\tau$ increases, the reduced time $\bar{t}$ decreases and the right-hand side of Eq.\,(\ref{ExpectationValueFinal}) increases.

What we have considered so far is a particle described by a wavefunction which is an eigenstate of the time-dependent harmonic oscillator Hamiltonian. It turns out, however, that Eq.\,(\ref{ChiSchrodinger}) also possesses a Gaussian wavepacket solution \cite{WavePackets} and, hence, a freely falling wavepacket can also be described based on the results we derived here. In fact, the relative change in time of the width $\braket{\tilde{x}^2}=\braket{x^2}-\braket{x}^2$ of the wavepacket is given by \cite{WavePackets},
\begin{equation}
\frac{1}{\braket{\tilde{x}^2}}\frac{{\rm d}\braket{\tilde{x}^2}}{{\rm d}\bar{t}}=\frac{2\dot{\gamma}(\bar{t})}{\gamma(\bar{t})}.
\end{equation}
Here, the function of time $\gamma(\bar{t})$ satisfies again an Ermakov differential equation of the form $\ddot{\gamma}+2\gamma/9\bar{t}^2=1/\gamma^3$. With the replacement $\gamma(\bar{t})\rightarrow\sqrt{m_i}\,\rho(\bar{t})$, this equation becomes identical to Eq.\,(\ref{RhoEquation}) which we already solved for $\rho(\bar{t})$. Therefore, using the results (\ref{Rho2FinalFinal}), we deduce that,
\begin{equation}\label{WidthChange}
\frac{1}{\braket{\tilde{x}^2}}\frac{{\rm d}\braket{\tilde{x}^2}}{{\rm d}\bar{t}}=\frac{12}{\bar{t}}\times\frac{3-5\left(\frac{\bar{t}}{\bar{t_0}}\right)^{-\frac{1}{3}}+2\left(\frac{\bar{t}}{\bar{t_0}}\right)^{-\frac{2}{3}}}{2+\left[5-6\left(\frac{\bar{t}}{\bar{t_0}}\right)^{-\frac{1}{3}}\right]^2}.
\end{equation}
This result gives the rate of shrinking of the width of the wavepacket as viewed from the freely-falling reference frame of the laboratory. Note that this rate of change of the width is proportional to $\dot{\gamma}(\bar{t})$, and hence also proportional to $\dot{\rho}(\bar{t})$. This means that an initial condition, for which $\dot{\rho}(\bar{t}_0)=0$, represents a vanishing initial shrinking of the wavepacket's width.

The remarkable feature of both results (\ref{ExpectationValueFinal}) and (\ref{WidthChange}) is that they are independent of the inertial mass $m_i$ of the particle. This would not have been possible to obtain if we had kept the distinction between inertial and gravitational masses. In fact, if we had done so, we would have had the ratio $m_g/m_i$ all over the place. Note that the mass-independence in these results is specifically due to the emergence of this ratio in an isolated form. The mass-independence of the expectation value of the Hamiltonian, as well as of the rate of change of the wavepacket's width is, therefore, what actually might more properly be viewed as the analog of the universality of free fall in quantum mechanics. Witnessing any experimental dependence of these results on the nature of the quantum objects used as test particles would signal an issue in our assumption of $m_i=m_g$. Note that these results are in agreement with the mass-independence that would emerge from Eq.\,(\ref{xyHarmonicEnergy}) of Sec.\,\ref{Sec:II} when one sets there $m_g=m_i$. The extra ingredient that has been gained here is the specific time-dependence of the energy of the particle and of the shape of a wavepacket, that would be witnessed within an {\it inertial} reference frame. This is a substitute for the $t^3$-phase effect discussed in the Introduction. Instead of measuring phase differences, one measures a time-dependence.

As a final remark in this section, we would like to point out that one could still use circular geodesics instead of the radial geodesic we have adopted here for the inertial laboratory. The metric that emerges then in the Fermi normal coordinates inside the inertial laboratory involves terms of the form $z^2\cos(2\Omega\tau)$ and $xz\sin(2\Omega\tau)$ \cite{Gabel}. Here, $\Omega=\sqrt{GM/r_0^3}$, with $r_0$ a radial coordinate of the revolving laboratory from the center of the Earth. The problem with such a case is that the time-dependent Schr\"odinger equation is rendered much harder to solve due to the wavefunction which becomes non-separable. The important fact to keep in mind, though, is that the cross term $xz$ leads again to a spatial entanglement of the wavefunction in the reference frame of the laboratory. Such experiments would, of course, need to be done on board of a satellite-based laboratory.
\section{Using the Gross-Pitaevskii equation}\label{Sec:IV}

For completeness, we are going to discuss briefly in this section the possibility of replacing the test-particle of the previous sections by a freely-falling Bose-Einstein condensate or a superfluid. However, as the latter require both analytic and numerical techniques that are completely different to the approach we adopted in this paper, it is clear that we cannot do justice to such a topic here. Nevertheless, we believe that the importance ---\! and the link\! --- of the topic to our present work deserves at least a short section to be devoted to a formal discussion of it in order to highlight the expected outcomes. 

Let us consider here a particle that obeys the nonlinear Gross-Pitaevskii equation, like what happens to the particles of a Bose-Einstein condensate \cite{GPE}. For that purpose, we need to assume, as often done in the literature (see, e.g., Refs,\,\cite{Gabel,Anandan,Suarez,KGIssue2}), the equation for the wavefunction to be the Gross-Pitaevskii equation in curved spacetime. Within the nonrelativistic approximations, the equation reduces to a nonlinear Schr\"odinger equation coupled to the Newtonian potential. As such, when taking into account tidal forces, the nonlinear equation we obtain for the case of a condensate freely-falling within the noninertial reference frame of the laboratory stuck to the ground is
\begin{equation}\label{GPE1}
-\frac{\hbar^2}{2m_i}\left(\partial_{x}^2+\partial_{y}^2+\partial_{\xi}^2\right)\psi+\frac{m_g}{2}\omega^2\left(x^2+y^2-2\xi^2\right)\psi\\+\frac{4\pi\hbar^2a_s}{m_i}|\psi|^2\psi=\mu\psi.
\end{equation}
Here, we have introduced the chemical potential $\mu$ inside which we included the constant term $\kappa$ defined below Eq.\,(\ref{RearrangedSchrodinger}) of Sec.\,\ref{Sec:II}. The constant $a_s$ represents the scattering length of the bosons of the condensate among themselves. This scattering gives rise either to a repulsive self-interaction of the bosons, for which $a_s>0$, or to an attractive self-interaction for which $a_s<0$. 

In the case of a condensate freely-falling inside a laboratory that is itself freely-falling, the gravitational potential to use is the one from Eq.\,(\ref{FermiNormalSchrodinger}). The Gross-Pitaevskii equation then reads
\begin{equation}\label{GPE2}
-\frac{\hbar^2}{2m_i}\left(\partial_{x}^2+\partial_{y}^2+\partial_z^2\right)\psi+\frac{m_g}{2}\omega^2(\tau)\left(x^2+y^2-2z^2\right)\psi\\+\frac{4\pi\hbar^2a_s}{m_i}|\psi|^2\psi=i\hbar\partial_\tau\psi.
\end{equation}
Here, the time-dependent angular frequency along the geodesic $r(\tau)$ is $\omega(\tau)=\sqrt{GM/r^3(\tau)}$. When both equations are solved, we expect that very interesting information could be extracted about the energy spectrum of the condensate, or the superfluid used as well as the time-evolution of its shape. However, given the peculiar potentials displayed in Eqs.\,(\ref{GPE1}) and (\ref{GPE2}), it is evident that even with the presently available numerical techniques for dealing with the nonlinear equation in the presence of harmonic potentials \cite{GPE1995A,GPE1995B,GPE1996A,GPE1996B,GPE1998,GPEReview2013}, one still needs to find new techniques to deal with our special cases.  In fact, what renders our cases a little bit different is the inverse harmonic oscillator. Given the nonlinearity of the equations (\ref{GPE1}) and (\ref{GPE2}), the separation of variables is not valid. Otherwise, we would have adopted a separate different numerical technique for each of the two kinds of harmonic oscillators. Nevertheless, according to what we found for the single-particle in the linear case, and according to what is already known for the anisotropic harmonic oscillator in the nonlinear case \cite{GPE1996A,GPE1996B,GPE1998}, we very much expect the emergence of a quantized energy-spectrum that would also depend on the isolated ratio $m_g/m_i$ of the particles making the condensate or the superfluid. As for the shape of the superfluid, numerical simulations could be used to extract specific behaviors that could be confronted with actual experiments. All this remains, unfortunately, beyond the scope of the present paper.   
\section{Summary and Discussion}
We have investigated the effects of tidal forces due to a spherically symmetric gravitational field on the free fall of a quantum particle ---\! or any quantum object for that matter\! --- from various different perspectives. We first examined the effects as they would be observed from a noninertial reference frame that would be stuck to the ground on the surface of the Earth. We found that, up to the second order in the nonrelativistic approximation, the resulting gravitational potential is that of a simple harmonic oscillator in the horizontal direction, but that in the vertical direction the potential is that of an inverted harmonic oscillator. We studied two different situations. One in which we imposed a fixed boundary condition on the wavefunction along the vertical direction, and the other in which we imposed time-varying boundary conditions along the vertical direction. The emerging pattern in each case is different. The dynamics of the particle in the first case does not give rise to an isolated ratio $m_g/m_i$, but in the second case it does as in classical mechanics.
The appearance of the isolated ratio $m_g/m_i$ in the second case was possible because inertial and gravitational effects share common contributions to the dynamics of the particle. This is, in fact, the reason why in the first case the ratio $m_g/m_i$ arises in an isolated form along the horizontal plane, but not in the vertical direction. It arises along the $xy$-plane thanks to the gravitationally induced harmonic oscillators. The harmonic oscillators along the $xy$-plane have been induced by the geometric shape of the gravitational source. As such, the geometric extension of the wavefunction, itself governed by the inertial properties of the particle, is the origin of the dependence of the energy spectrum (\ref{xyHarmonicEnergy}) on the pure ratio $m_g/m_i$.

Still, this was achieved at the price of introducing the gravitational mass concept which has only a meaning within Newtonian gravity. 
In fact, as was already briefly mentioned below Eq.\,(\ref{KGCST}), general relativity does not allow a distinction between a gravitational mass and an inertial mass \footnote{Note that we solely focus here on general relativity and its foundations, excluding, thus, from our discussion extensions of general relativity.  For a discussion on the status of the equivalence principle in, for example, the so-called dilaton models, see, e.g., Refs.\,\cite{R11,R12}, as well as Refs.\,\cite{R2,R3}.}. The reason why we introduced such a distinction in our treatment is to be able to investigate possible departures from general relativity's foundations at the quantum level. The mere fact of using the Klein-Gordon equation in curved spacetime is itself an implicit assumption of the equivalence between the two masses. Nevertheless, introducing by hand a different concept for the mass is one way for testing how fundamental is the identity between the two masses. In fact, testing the universality of free fall will probably not going to confirm or disproof the geometrization of gravity at the quantum level, and hence might not be capable of taking us closer to the theory of quantum gravity when the latter is viewed as a quantization of spacetime, as already argued in Ref.\,\cite{Philosophy}. However, we believe that investigating the various ways gravity and inertia enter into the dynamics of quantum particles might at least help us understand better the nature of gravity and the way it influences quantum objects. 

One of our goals behind introducing the gravitational mass $m_g$ was in the hope of being one day able to detect any possible difference between the two masses. Our main goal, however, remains to use $m_g$ as a tool to probe the role the different concepts of mass play when coupling quantum objects to gravity. Actually, the correctness of general relativity is, in our opinion, sufficient to take the equivalence between the two masses to be an unequivocal fact about Nature. This equivalence can even be classified in the same category as the constancy of the speed of light. While the constancy of the speed of light can be deduced from the fundamental principle that the laws of Nature should be the same in all inertial reference frames, the equivalence of the two masses can be viewed as a consequence of the fact that gravity is equivalent to geometry. The latter equivalence is, of course, itself a consequence of extending the invariance of the laws of Nature to arbitrary reference frames. The status of the identity between the two concepts of mass is therefore reminiscent of the status of the constancy of the speed of light towards the end of the 19th century. The geometrization of gravity is to the equivalence of the two masses what the invariance of the laws of Nature is to the constancy of the speed of light. Thereby, taking the trouble to formally distinguish between the two concepts of mass at the quantum level can certainly take us one step further towards a better understanding of the equivalence of the two masses ({\it i.e.} of the emergence of geometry) at the quantum level.

By studying the free fall of a particle in a reference frame which is itself free-falling, our work showed that assuming the two masses are equivalent leads to the neat result that the effect of tidal forces induces a simple harmonic motion on the particle in which the latter acquires a mass-independent and quantized energy. In addition, the acquired energy is time-dependent. Furthermore, we found that when a wavepacket is considered, the relative change in the width of the latter is also mass-independent. We argued that these effects are to be seen as the analogs of the universality of free fall in classical mechanics. Nevertheless, any experimental deviation could be taken to be a possible sign for the violation of the assumption that the two concepts of mass should be identified. In fact, our primary motivation in this work is based on the belief that not finding a departure from the equivalence principle at the classical level begs for a deeper investigation of why no possible departure from it could be found so far at the quantum level as well.

We have to emphasize here the challenges that would be encountered in any attempt to conduct an experimental investigation of our results. The first major hindrance will, of course, be encountered due to the weakness of the effects we derived here. We saw in Sec.\,\ref{Sec:II.a} that the energy separation of the levels of the gravitationally-induced harmonic oscillators is of the order of $\Delta E=\hbar\sqrt{g/R_E}\sim10^{-18}\,$eV. This order of magnitude is not improved by using the tidal forces offered by smaller-radius bodies like the Moon. In fact, the value one gets on the Moon for $\Delta E$ is very close to this one, for even though the radius of the Moon is smaller the gravitational acceleration $g$ on the latter is smaller too. The second hindrance one seems to encounter experimentally is the fact that Earth is not homogeneous and not perfectly spherical as we assumed here. However, the corrections one needs to bring due to this departure from sphericity are not important as long as the $1/r$-potential stays a good approximation. This is indeed the case, for the corrections brought to such a potential scale like $1/r^3$. Actually, one might still extract a better approximation for the potentials created at an external point by an Earth assumed to be an ellipsoid instead of a perfect sphere. Unfortunately, working with ellipsoidal and nonhomogeneous distributions of mass is too involved (see, e.g., Ref.\,\cite{Chandrasekhar}). The other experimental challenge one would be confronted to is the smallness of the inverse time constant for the decay rate of the quantized energy of the system, $\omega=\sqrt{g/R_E}\approx10^{-3}\,$s$^{-1}$, in Sec.\,\ref{Sec:II.b}. This entails that one has to be able to sustain quantum coherence up to two hours. Finally, another major experimental challenge we saw is that the length of the free fall time required in Sec.\,\ref{Sec:III} has to be at least of the order of $38$ seconds even for the unpractical case of very large initial speeds.

We expect that the experimental challenges we just listed could be overcome with more technological advances. The good news, however, is that the only result that relies on the gravitational acceleration $g$, and hence requires a large gravitational source, is the energy spectrum given in Eq.\,(\ref{EtaDiscretization}). That result requires indeed, like the phase shift (\ref{COWPhase}) and the energy spectrum (\ref{AiryQuantization}), to have a large $g$. All our other results depend either (i) on $\omega$, like in Eqs.\,(\ref{xyHarmonicEnergy}) and (\ref{Hamiltonian}) or (ii) on $\bar{t}_0$, like in Eqs.\,(\ref{ExpectationValueFinal}) and (\ref{WidthChange}). Now, we know that for any solid sphere of mass density $\rho$ and radius $R$, our $\omega$ can be written as $\sqrt{g/R}=\sqrt{4\pi G\rho}$. This means that for the results (\ref{xyHarmonicEnergy}) and (\ref{Hamiltonian}) the radius of the mass source is not important. What is important is the mass density of the material used. As such, a material of very high mass density, like platinum or osmium for which the density is about $\sim22\,$g/cm$^3$, could be exploited. In fact, with such a density one easily reaches an $\omega$ of the order of $4\times10^{-3}\,$s$^{-1}$. On the other hand, we also have $\bar{t}_0=\sqrt{2r_0^3/9GM}=\sqrt{r_0^3/(18\pi G\rho R^3)}$ in Eqs.\,(\ref{ExpectationValueFinal}) and (\ref{WidthChange}), so that for such a density and for an initial separation $r_0$ of the test particle from the center of the sphere not much different from $R$, we have $\bar{t}_0\sim110$ seconds. Therefore, we see that one can still experimentally benefit from the possibility of using in the lab a homogeneous spherically symmetric massive material, instead of relying on the gravitational field of the Earth to detect tidal forces.

Our main goal in this paper has been, as in Ref.\,\cite{ConformalKG}, to achieve a better understanding of the interaction of gravity with quantum particles. We have mentioned that gravity might be seen as interacting with a quantum particle via the wavefunction of the latter. It follows that the universality of gravity can thus be understood as due to the universality of quantum mechanics. However, what we mean here by the word universality is not the universality of the free fall of quantum objects, but rather the universality of the {\it coupling} of gravity to mass and energy. In other words, by universality we mean that all objects couple to gravity via their inertial properties, instead of a new hypothetical gravitational charge. In fact, as we discussed below Eq.\,(\ref{KGCST}), when looking at the Klein-Gordon equation in curved spacetime, it appears as if gravity is acting on the wavefunction of the particle rather than acting directly on the latter. This observation is, in itself, very interesting despite the big debate in literature around the reality of the wavefunction (see, e.g., Ref.\,\cite{Reality}). As such, we may very well conclude that it is the universality of quantum mechanics that actually guarantees the universality of gravity in the sense we gave here to the word universality. Therefore, it is not excluded that understanding quantum mechanics would automatically lead to the emergence of the concept of gravity.
\appendix
\section{Solving the Ermakov equation (\ref{RhoEquation})}\label{A}
We follow here the method (outlined in Ref.\,\cite{FormulasBook1}, Sec.\,2.9.1.2) to solve the Ermakov (sometimes also called Yermakov) equation (\ref{RhoEquation}) and give the detailed steps that lead us to its solution. The general Ermakov equation has the form \cite{FormulasBook1},
\begin{equation}\label{GeneralErmakov}
y''+f(x)y=\frac{a}{y^3},
\end{equation}
where a prime denotes here a derivative with respect to the variable $x$. In our case, Eq.\,(\ref{RhoEquation}) is of this form, but with the variable $x$ replaced by the variable $\bar{t}$, the function $f(x)$ is replaced by the function $2/9\bar{t}^2$, and the constant $a$ is replaced by $1/m_i^2$. The first step for solving Eq.\,(\ref{RhoEquation}) consists then in finding a function of time $w(\bar{t})$ which is a solution of the second-order linear differential equation,
\begin{equation}\label{wEquation}
\ddot{w}+\frac{2}{9\bar{t}^2}w=0.
\end{equation}
With the change of variables $\eta=1/\bar{t}$, this equation becomes the second-order Euler's differential equation in the variable $\eta$, the solution of which is given in Sec.\,2.1.2.123 of Ref.\,\cite{FormulasBook1}. Using that solution, we find,
\begin{equation}\label{wSolution}
w(\bar{t})=A_1\bar{t}^\frac{2}{3}+A_2\bar{t}^\frac{1}{3}.
\end{equation}
where, $A_1$ and $A_2$ are two integration constants. Note, however, that keeping both terms in this solution would make our subsequent analysis extremely cumbersome without leading to easily trackable physical insights about the system. Fortunately, we can appeal to the following powerful simple argument which turns out to be very useful as a guide for solving our Ermakov equation (\ref{RhoEquation}) and extracting valuable physical insights. 

We note that the crucial differential equation (\ref{wEquation}), which is a necessary intermediate step for solving our equation (\ref{RhoEquation}), is time-reversal symmetric. We therefore set the constant $A_2$ to zero in Eq.\,(\ref{wSolution}) so that the solution $w(\bar{t})$ becomes invariant under time reversal as well. This would, indeed, keep Eq.\,(\ref{wEquation}) time-reversal symmetric without having to introduce complex numbers. In fact, if we keep the second term in Eq.\,(\ref{wSolution}), a time-reversal symmetry of Eq.\,(\ref{wEquation}) would require performing a complex phase redefinition of the integration constant $A_2$ in the solution (\ref{wSolution}). This simple argument would not have been possible, however, if we had kept $m_g\neq m_i$ in Eq.\,(\ref{FermiNormalSchrodinger}), as we would then have had to work with an even more involved expression than Eq.\,(\ref{wSolution}). But, once again, if we had done so our equations would only be rendered more cumbersome without gaining new insights about the time dependence of the system. Let us now, in addition, redefine the constant $A_1$ in terms of a new constant $A$ and a time scale $\bar{t}_0$, so that the physical dimensions of $w(\bar{t})$ become entirely given by those of the constant $A$, which, in turn, becomes easier to find as we will see shortly. The solution we take then for $w(\bar{t})$ is of the form $w(\bar{t})=A\left(\bar{t}/\bar{t}_0\right)^\frac{2}{3}$. This turns out to be a very convenient form as we will see below.

The second and last step consists now in evaluating an integral involving $w(\bar{t})$ \cite{FormulasBook1}. The general solution of Eq.\,(\ref{RhoEquation}) is indeed given by,
\begin{align}\label{RhoSolutionAppendix}
C_1\rho^2(\bar{t})&=\frac{w^2(\bar{t})}{m_i^2}+w^2(\bar{t})\left(C_2+C_1\int\frac{{\rm d}\bar{t}}{w^2(\bar{t})}\right)^2\nonumber\\
&=A^2\left(\frac{\bar{t}}{\bar{t_0}}\right)^\frac{4}{3}\left(\frac{1}{m_i^2}+\left[C_2-\frac{3\bar{t}_0C_1}{A^2}\left(\frac{\bar{t}}{\bar{t_0}}\right)^{-\frac{1}{3}}+C_3\right]^2\right),
\end{align}
where $C_1$, $C_2$ and $C_3$ are all integration constants. After dividing both sides of this equation by the constant $C_1$, it becomes evident that we may absorb $C_1$ into the constant $A^2$. Next, we combine the constants $C_2$ and $C_3$ inside the square brackets into a single constant $B$. The presence of the term $1/m_i^2$ inside the parentheses indicates that the constant $B$ should have the physical dimensions of $1/m_i$. Let us therefore introduce the dimensionless constant $\alpha$ such that $B=\alpha/m_i$. We obtain then the following expression for $\rho(\bar{t})$:
\begin{equation}\label{RhoIntermediate}
\rho^2(\bar{t})=\frac{A^2}{m_i^2}\left(\frac{\bar{t}}{\bar{t_0}}\right)^\frac{4}{3}+\frac{A^2}{m_i^2}\left[\alpha \left(\frac{\bar{t}}{\bar{t_0}}\right)^\frac{2}{3}-\frac{3m_i\bar{t}_0}{A^2}\left(\frac{\bar{t}}{\bar{t_0}}\right)^\frac{1}{3}\right]^2.
\end{equation}
From this expression, we immediately deduce the physical dimensions of the constant $A$. In fact, from Eq.\,(\ref{RhoEquation}), we know that the physical dimensions of $\rho^2(\bar{t})$ should be time over mass. Therefore, we conclude that the physical dimensions of $A^2$ are time times mass. In other words, $A^2$ has the same physical dimensions as those of $\hbar/c^2$. Let us then set the constant $A^2$ to be proportional to the physical constants already involved in the system: $m_i$ and $\bar{t}_0$. Setting $A^2=\beta\,m_i\bar{t}_0$ for some arbitrary dimensionless constant $\beta$, expression (\ref{RhoIntermediate}) yields,
\begin{align}\label{Rho2Final}
\rho^2(\bar{t})&=\frac{\beta\,\bar{t}_0}{m_i}\left(\frac{\bar{t}}{\bar{t_0}}\right)^\frac{4}{3}+\frac{\beta\,\bar{t}_0}{m_i}\left[\alpha\left(\frac{\bar{t}}{\bar{t_0}}\right)^\frac{2}{3}-\frac{3}{\beta}\left(\frac{\bar{t}}{\bar{t_0}}\right)^\frac{1}{3}\right]^2,\nonumber\\
\dot{\rho}^2(\bar{t})&=\frac{4\beta}{9m_i\bar{t}_0}\left(\frac{\bar{t}_0}{\bar{t}}\right)^\frac{2}{3}\left[1+\alpha^2-\frac{9\alpha}{2\beta}\left(\frac{\bar{t}}{\bar{t_0}}\right)^{-\frac{1}{3}}+\frac{9}{2\beta^2}\left(\frac{\bar{t}}{\bar{t_0}}\right)^{-\frac{2}{3}}\right]^2\left(1+\left[\alpha-\frac{3}{\beta}\left(\frac{\bar{t}}{\bar{t_0}}\right)^{-\frac{1}{3}}\right]^2\right)^{-1}.
\end{align}
Note that from the first identity, we deduce that $\beta$ has to be a positive constant. In order to fix the arbitrary constants $\alpha$ and $\beta$, we evaluate $\rho^2(\bar{t})$ and $\dot{\rho}^2(\bar{t})$ at $\bar{t}=\bar{t}_0$, {\it i.e.}, for $\tau=0$. We find,
\begin{align}\label{RhoDotRhoInitial}
\rho^2(\bar{t}_0)&=\frac{\beta\,\bar{t}_0}{m_i}\left[1+\left(\alpha-\frac{3}{\beta}\right)^2\right],\nonumber\\
\dot{\rho}^2(\bar{t}_0)&=\frac{4\beta}{9m_i\bar{t}_0}\left(1+\alpha^2-\frac{9\alpha}{2\beta}+\frac{9}{2\beta^2}\right)\!\left[1+\left(\alpha-\frac{3}{\beta}\right)^2\right]^{-1}\!\!.
\end{align}
Setting now $\alpha=\frac{1}{4\beta}\left(9\pm\sqrt{9-16\beta^2}\right)$, leads to the initial condition $\dot{\rho}^2(\bar{t}_0)=0$, The meaning of this condition is elucidated below Eq.\,(\ref{WidthChange}). On the other hand, for such an expression of $\alpha$, we have
\begin{equation}\label{Rho2Initial}
\rho^2(\bar{t}_0)=\frac{\bar{t}_0}{m_i}\left[\beta+\frac{9}{16\beta}\left(1\pm\sqrt{1-\frac{16}{9}\beta^2}\right)^2\right].
\end{equation}
By plugging these initial expressions into the expectation value (\ref{ExpectationValue}), and requiring that the latter to be identical to the initial expectation value of the corresponding time-independent harmonic oscillator, we can easily fix the constant of integration $\beta$ and then deduce the integration constant $\alpha$ as well. 

Indeed, inserting expression (\ref{Rho2Initial}) as well as $\dot{\rho}(\bar{t}_0)=0$ into Eq.\,(\ref{ExpectationValue}), we find the following initial-time expectation value:
\begin{equation}\label{ExpectationValueAppendix}
\!\!\bra{n}{H_x}\ket{n}_{\bar{t}=\bar{t}_0}=\frac{\hbar\left(n+\frac{1}{2}\right)}{2\bar{t}_0}\left(\left[\frac{2\beta}{9}+\frac{1}{8\beta}\left(1\pm\sqrt{1-\frac{16}{9}\beta^2}\right)^2\right]\right)\!\left[\beta+\frac{9}{16\beta}\left(1\!\pm\!\sqrt{1-\frac{16}{9}\beta^2}\right)^2\right]^{-1}\left.\vphantom{\left[\frac{2\beta}{9}+\frac{1}{8\beta}\left(1\!\pm\!\sqrt{1-\frac{16}{9}\beta^2}\right)^{2}\right]}\!\right)\!\!.
\end{equation}
On the other hand, according to Eq.\,(\ref{ChiSchrodinger}) the initial angular frequency $\omega(\bar{t})$ of the time-dependent harmonic oscillator is $\omega_0=\sqrt{2}/3\bar{t}_0$. The expectation value of the corresponding time-independent harmonic oscillator is then $\hbar\omega_0\left(n+\frac{1}{2}\right)=\hbar\sqrt{2}\left(n+\frac{1}{2}\right)/3\bar{t}_0$. The initial expectation value (\ref{ExpectationValueAppendix}) would coincide with this value if and only if we set the dimensionless constant $\beta$ to be $1/\sqrt{2}$. This real value for $\beta$ results only when picking up the plus sign in front of the square roots in Eq.\,(\ref{ExpectationValueAppendix}). The minus sign does not lead to any real value for $\beta$. Therefore, we have the following final values for $\alpha$ and $\beta$:
\begin{equation}\label{AlphaBeta}
\alpha=\frac{5\sqrt{2}}{2},\qquad \beta=\frac{\sqrt{2}}{2}.
\end{equation}
Substituting these into expressions (\ref{Rho2Final}), we get the final expressions for $\rho^2(\bar{t})$ and $\dot{\rho}^2(\bar{t})$: 
\begin{align}\label{Rho2FinalFinal}
\rho^2(\bar{t})&=\frac{\sqrt{2}\,\bar{t}_0}{4m_i}\left(\frac{\bar{t}}{\bar{t_0}}\right)^\frac{4}{3}\left(2+\left[5-6\left(\frac{\bar{t}}{\bar{t_0}}\right)^{-\frac{1}{3}}\right]^2\right),\nonumber\\
\dot{\rho}^2(\bar{t})&=\frac{9\sqrt{2}}{m_i\bar{t}_0}\left(\frac{\bar{t}_0}{\bar{t}}\right)^{\frac{2}{3}}\left[3-5\left(\frac{\bar{t}}{\bar{t_0}}\right)^{-\frac{1}{3}}+2\left(\frac{\bar{t}}{\bar{t_0}}\right)^{-\frac{2}{3}}\right]^2\nonumber\left(2+\left[5-6\left(\frac{\bar{t}}{\bar{t_0}}\right)^{-\frac{1}{3}}\right]^2\right)^{-1}.\\
\end{align}
\section*{Acknowledgments}
This work is supported in part by the SRC Interdisciplinary Team Grant from Bishop's University and by the Natural Sciences and Engineering Research Council of Canada (NSERC) Discovery Grant No. RGPIN-2017-05388.


\begin{thebibliography}{}
\bibitem{Sonego1995} S. Sonego, ``Is there a spacetime geometry?'', \href{https://www.sciencedirect.com/science/article/abs/pii/037596019500743M}{Phys. Lett. A {\bf208}, 1 (1995)}.

\bibitem{Lammerzahl1996} C. L\"ammerzahl, ``On the equivalence principle in quantum theory'', \href{https://link.springer.com/article/10.1007/BF02113157}{Gen. Rel. Grav. {\bf28}, 1043 (1996)} [\href{https://arxiv.org/abs/gr-qc/9605065}{arXiv:gr-qc/9605065}].

\bibitem{Lorenza} L. Viola and R. Onofrio, ``Testing the equivalence principle through freely falling quantum objects'', \href{https://journals.aps.org/prd/abstract/10.1103/PhysRevD.55.455}{Phys. Rev. D{\bf55}, 455 (1997)} [\href{https://arxiv.org/abs/quant-ph/9612039}{arXiv:quant-ph/9612039}].

\bibitem{Davies0} P.\,C.\,W. Davies, ``Quantum mechanics and the equivalence principle'', \href{https://iopscience.iop.org/article/10.1088/0264-9381/21/11/017}{Class. Quantum Grav. {\bf21}, 2761 (2004)} [\href{https://arxiv.org/abs/quant-ph/0403027}{arXiv:quant-ph/0403027}].

\bibitem{Philosophy} E. Okon and C. Callender, ``Does Quantum Mechanics Clash with the Equivalence Principle - and Does it Matter?'', \href{https://link.springer.com/article/10.1007/s13194-010-0009-z}{Euro. Jnl. Phil. Sci. {\bf1}, 133 (2011)} [\href{https://arxiv.org/abs/1008.5192}{arXiv:1008.5192}].

\bibitem{NonEquivalenceEP} E. Di Casola, S. Liberati, S. Sonego, ``Nonequivalence of equivalence principles'', \href{https://aapt.scitation.org/doi/10.1119/1.4895342}{Am. J. Phys. {\bf83}, 39 (2015)} [\href{https://arxiv.org/abs/1310.7426v2}{arXiv:1310.7426v2}].

\bibitem{AmjP} M. Nauenberg, ``Einstein's equivalence principle in quantum mechanics revisited'', \href{https://aapt.scitation.org/doi/full/10.1119/1.4962981}{Am. J. Phys. {\bf84}, 879 (2016)}.

\bibitem{Nature} M. Zych and C. Brukner, ``Quantum formulation of the Einstein Equivalence Principle'', \href{https://www.nature.com/articles/s41567-018-0197-6}{Nature Phys. {\bf14}, 1027 (2018)} [\href{https://arxiv.org/abs/1502.00971}{arXiv:1502.00971}].

\bibitem{WEP2019} P.\,C.\,M. Flores and E.\,A. Galapon, ``Quantum free fall motion and quantum violation of weak equivalence principle'', \href{https://journals.aps.org/pra/abstract/10.1103/PhysRevA.99.042113}{Phys. Rev. A {\bf99}, 042113 (2019)} [\href{https://arxiv.org/abs/1808.02646}{arXiv:1808.02646}].

\bibitem{Will2018} C.\,M. Will, {\it Theory and Experiment in Gravitational Physics}, 2nd Edition (Cambridge University Press, Cambridge, 2018).
\bibitem{Speliotopoulos} A.\,D. Speliotopoulos and R.\,Y. Chiao, ``Coupling of Linearized Gravity to Nonrelativistic Test Particles: Dynamics in the General Laboratory Frame'', \href{https://journals.aps.org/prd/abstract/10.1103/PhysRevD.69.084013}{Phys. Rev. D{\bf69}, 084013 (2004)} [\href{https://arxiv.org/abs/gr-qc/0302045}{arXiv:gr-qc/0302045}].

\bibitem{Davies} P.\,C.\,W. Davies, ``Transit time of a freely falling quantum particle in a background gravitational field'', \href{https://iopscience.iop.org/article/10.1088/0264-9381/21/24/001/meta}{Class. Quantum Grav. {\bf21}, 5677 (2004)} [\href{https://arxiv.org/abs/quant-ph/0407028}{arXiv:quant-ph/0407028}].

\bibitem{R3} A. Roura, ``Gravitational Redshift in Quantum-Clock Interferometry'', \href{https://journals.aps.org/prx/abstract/10.1103/PhysRevX.10.021014}{Phys. Rev. X {\bf10}, 021014 (2020)}.

\bibitem{R4} J. Audretsch and K-P. Marzlin, ``Ramsey fringes in atomic interferometry: Measurability of the influence of space-time curvature'', \href{https://journals.aps.org/pra/abstract/10.1103/PhysRevA.50.2080}{Phys. Rev. A {\bf50}, 2080 (1994)}.

\bibitem{R5} J. Audretsch and K-P. Marzlin, ``Atom interferometry with arbitrary laser configurations : exact phase shift for potentials including inertia and gravitation'', \href{https://jp2.journaldephysique.org/articles/jp2/abs/1994/11/jp2v4p2073/jp2v4p2073.html}{J. Phys. II France {\bf4}, 2073 (1994)}.

\bibitem{R6} P. Asenbaum, ``Phase shift in atom interferometry due to spacetime curvature'', \href{https://journals.aps.org/prl/abstract/10.1103/PhysRevLett.118.183602}{Phys. Rev. Lett. {\bf118}, 183602 (2017)} [\href{https://arxiv.org/abs/1610.03832}{arXiv:1610.03832}].

\bibitem{R7} A. Roura, ``Circumventing Heisenberg’s Uncertainty Principle in Atom Interferometry Tests of the Equivalence Principle'', \href{https://journals.aps.org/prl/abstract/10.1103/PhysRevLett.118.160401}{Phys. Rev. Lett. {\bf118}, 160401 (2017)} [\href{https://arxiv.org/abs/1509.08098}{arXiv:1509.08098}].

\bibitem{R8} C. Overstreet {\it el al}., ``Effective inertial frame in an atom interferometric test of the equivalence principle'', \href{https://journals.aps.org/prl/abstract/10.1103/PhysRevLett.120.183604}{Phys. Rev. Lett. {\bf120}, 183604 (2018)} [\href{https://arxiv.org/abs/1711.09986}{arXiv:1711.09986}].

\bibitem{R10} C. Anastopoulos, B.\,L. Hu, ``Equivalence principle for quantum systems: dephasing and phase shift of free-falling particles'', \href{https://iopscience.iop.org/article/10.1088/1361-6382/aaa0e8/meta}{Class. Quantum Grav. {\bf35}, 035011 (2018)} [\href{https://arxiv.org/abs/1707.04526}{arXiv:1707.04526}].

\bibitem{QuantumGalileo} Md. Manirul Ali {\it et al}., ``On the quantum analogue of Galileo's leaning tower experiment'', \href{https://iopscience.iop.org/article/10.1088/0264-9381/23/22/024}{Class. Quant. Grav. {\bf23}, 6493 (2006)} [\href{https://arxiv.org/abs/quant-ph/0606183}{arXiv:quant-ph/0606183}].

\bibitem{NonCommutativeQM} K.\,H.\,C. Castello-Branco and A.\,G. Martins, ``Free-fall in a uniform gravitational field in non-commutative quantum mechanics'', \href{https://aip.scitation.org/doi/10.1063/1.3466812}{J. Math. Phys. {\bf51}, 102106 (2010)} [\href{https://arxiv.org/abs/0803.0981}{arXiv:0803.0981}].

\bibitem{R2} D. Schlippert {\it et al}., ``Quantum Test of the Universality of Free Fall'', \href{https://journals.aps.org/prl/abstract/10.1103/PhysRevLett.112.203002}{Phys. Rev. Lett. {\bf112}, 203002 (2014)} [\href{https://arxiv.org/abs/1406.4979}{arXiv:1406.4979}].

\bibitem{AntiMatterFF2015} G. Dufour {\it et al}., ``Prospects for Studies of the Free Fall and Gravitational Quantum States of Antimatter'', \href{https://www.hindawi.com/journals/ahep/2015/379642/}{Adv. High E. Phys. {\bf2015}, 379642, (2015)}.

\bibitem{EP2017} L. Seveso, V. Peri and M.\,G.\,A. Paris, ``Does universality of free-fall apply to the motion of quantum probes?'', \href{https://iopscience.iop.org/article/10.1088/1742-6596/880/1/012067}{J. Phys.: Conf. Ser. {\bf880}, 012067 (2017)} [\href{https://arxiv.org/abs/1702.07526}{arXiv:1702.07526}].

\bibitem{R1} P. Asenbaum {\it et al}., ``Atom-interferometric test of the equivalence principle at the \begin{math} 10^{-12}\end{math} level'', \href{https://journals.aps.org/prl/abstract/10.1103/PhysRevLett.125.191101}{Phys. Rev. Lett. {\bf125}, 191101 (2020)} [\href{https://arxiv.org/abs/2005.11624}{arXiv:2005.11624}].

\bibitem{Colcelli(2020)} A. Colcelli {\it et al}., ``Free Fall of a Quantum Many-Body System'', \href{https://arxiv.org/abs/2009.03744}{arXiv:2009.03744}.
\bibitem{QBounce1971} P.\,W. Langhoff, ``Schr\"odinger Particle in a Gravitational Well'', \href{https://aapt.scitation.org/doi/10.1119/1.1986333}{Am. J. Phys. {\bf39}, 954 (1971)}.

\bibitem{COW1974} A.\,W. Overhauser and R. Collela, ``Experimental test of gravitational induced quantum interference'', \href{https://journals.aps.org/prl/abstract/10.1103/PhysRevLett.33.1237}{Phys. Rev. Lett. {\bf33}, 1237 (1974)}

\bibitem{COW1975} R. Colella, A.\,W. Overhauser, and S.\,A. Werner, ``Observation of gravitationally induced quantum interference'', \href{https://journals.aps.org/prl/abstract/10.1103/PhysRevLett.34.1472}{Phys. Rev. Lett. {\bf34}, 1472 (1975)}.

\bibitem{COWReview} D.\,M. Greenberger and A.\,W. Overhauser, ``Coherence effects in neutron diffraction and gravity experiments'', \href{https://journals.aps.org/rmp/abstract/10.1103/RevModPhys.51.43}{Rev. Mod. Phys. {\bf51} 43 (1979)}.

\bibitem{DroppingAtoms} A. Peters, K.\,Y. Chung and S. Chu, ``Measurement of gravitational acceleration by dropping atoms'', \href{https://www.nature.com/articles/23655#citeas}{Nature {\bf400}, 849 (1999)}.
\bibitem{QBounceNature1} V.\,V. Nesvizhevsky {\it et al.}, ``Quantum states of neutrons in the Earth's gravitational field'', \href{https://www.nature.com/articles/415297a}{Nature {\bf415}, 297 (2002)}.

\bibitem{QBounceNature2} T.\,J. Bowles, ``Quantum effects of gravity'', \href{https://www.nature.com/articles/415267a}{Nature {\bf415}, 267 (2002)}. 

\bibitem{QBouncePhysToday2002} B. Schwarzschild, ``Ultracold Neutrons Exhibit Quantum States in the Earth's Gravitational Field'', \href{https://physicstoday.scitation.org/doi/abs/10.1063/1.1472382}{Physics Today {\bf55}, 20 (2002)}. 

\bibitem{R9} E. Kajari {\it et al.}, ``Inertial and gravitational mass in quantum mechanics'', \href{https://link.springer.com/article/10.1007/s00340-010-4085-8}{Appl. Phys. B {\bf100}, 43 (2010)} [\href{https://arxiv.org/abs/1006.1988}{arXiv:1006.1988}]. 


\bibitem{QBounce2003} V.\,V. Nesvizhevsky {\it et al.}, ``Measurement of quantum states of neutrons in the Earth's gravitational field'', \href{https://journals.aps.org/prd/abstract/10.1103/PhysRevD.67.102002}{Phys. Rev. D {\bf67}, 102002 (2003)} [\href{https://arxiv.org/abs/hep-ph/0306198}{arXiv:hep-ph/0306198}].

\bibitem{Landry} A. Landry and M.\,B. Paranjape, ``Gravitationally induced quantum transitions", \href{https://journals.aps.org/prd/abstract/10.1103/PhysRevD.93.122006}{Phys. Rev. D {\bf93}, 122006 (2016)}
[\href{https://arxiv.org/abs/1601.06132}{arXiv:1601.06132}].

\bibitem{FFReview2017} M.S. Safronova {\it et al.}, ``Search for New Physics with Atoms and Molecules'', \href{https://journals.aps.org/rmp/abstract/10.1103/RevModPhys.90.025008}{Rev. Mod. Phys. {\bf90}, 025008 (2018)} [\href{https://arxiv.org/abs/1710.01833}{arXiv:1710.01833}].

\bibitem{Science} V. Xu {\it et al.}, ``Probing gravity by holding atoms for 20 seconds",
 \href{https://science.sciencemag.org/content/366/6466/745}{Science {\bf366}, 745 (2019)} [\href{https://arxiv.org/abs/1907.03054}{arXiv:1907.03054}].
 
 \bibitem{GravityLandauI} A. Landry and F. Hammad, ``Landau levels in a gravitational field: The Schwarzschild spacetime case'', \href{https://www.mdpi.com/2218-1997/7/5/144}{Universe {\bf7}(5), 144 (2021)} [\href{https://arxiv.org/abs/1909.01827}{arXiv:1909.01827}].

\bibitem{GravityLandauII} F. Hammad and A. Landry, ``Landau levels in a gravitational field: The Levi-Civita and Kerr spacetimes case'', \href{https://link.springer.com/article/10.1140%2Fepjp%2Fs13360-020-00108-1}{Eur. Phys. J. Plus {\bf135} (2020) 90} [\href{https://arxiv.org/abs/1910.01899}{arXiv:1910.01899v2}].

\bibitem{Josephson} F. Hammad and A. Landry, ``A simple superconducting quantum interference device for testing gravity'', \href{https://www.worldscientific.com/doi/abs/10.1142/S0217732320501710}{Mod. Phys. Lett. A {\bf35} (2020) 2050171} [\href{https://arxiv.org/abs/2005.05798}{arXiv:2005.05798}].

\bibitem{COWBall} F. Hammad, A. Landry and K. Mathieu, ``Prospects for testing the inverse-square law and gravitomagnetism using quantum interference",
 \href{https://www.worldscientific.com/doi/abs/10.1142/S0218271821500048}{Int. J. Mod. Phys. D. {\bf30}, 2150004 (2021)} [\href{https://arxiv.org/abs/1910.13814}{arXiv:1910.13814}].
 
\bibitem{QHE} F. Hammad, A. Landry and K. Mathieu, ``A fresh look at the influence of gravity on the quantum Hall effect'', \href{https://link.springer.com/article/10.1140/epjp/s13360-020-00481-x}{Eur. Phys. J. Plus {\bf135} 449 (2020)} [\href{https://arxiv.org/abs/2005.10631}{arXiv:2005.10631}].
\bibitem{Gabel} O. Gabel, ``Bose-Einstein Condensates in Curved Space-Time -- From Concepts of General Relativity to Tidal Corrections for Quantum Gases in Local Frames'', (PhD. Thesis, \href{http://tuprints.ulb.tu-darmstadt.de/8809/}{Darmstadt, Technische Universit\"at, (2019)}).
\bibitem{t3Phase} C. Marletto and V. Vedral, ``On the testability of the equivalence principle as a gauge principle detecting the gravitational t$^3$ phase'', \href{https://www.frontiersin.org/articles/10.3389/fphy.2020.00176/full}{Front. Phys. {\bf8}, 176 (2020)} [\href{https://arxiv.org/abs/2004.11616}{arXiv:2004.11616}].
\bibitem{Catalog} T. Mueller and F. Grave, ``Catalogue of Spacetimes'', \href{https://arxiv.org/abs/0904.4184}{arXiv:0904.4184}.
\bibitem{SchroederAjP} D.\,V. Schroeder, ``Entanglement isn't just for spin'', \href{https://aapt.scitation.org/doi/10.1119/1.5003808}{Am. J. Phys. {\bf85}, 812 (2017)} [\href{https://arxiv.org/abs/1703.10620}{arXiv:1703.10620}].

\bibitem{EPRinCST} H. Terashima and M. Ueda, ``Einstein-Podolsky-Rosen correlation in gravitational field'', \href{https://journals.aps.org/pra/abstract/10.1103/PhysRevA.69.032113}{Phys. Rev. A{\bf69}, 032113 (2004)} [\href{https://arxiv.org/abs/quant-ph/0307114}{arXiv:quant-ph/0307114}].

\bibitem{BruschiNPj} D.\,E. Bruschi {\it et al}., ``Testing the effects of gravity and motion on quantum entanglement in space-based experiments'', \href{https://iopscience.iop.org/article/10.1088/1367-2630/16/5/053041}{New J. Phys. {\bf16}, 053041 (2014)} [\href{https://arxiv.org/abs/1306.1933}{arXiv:1306.1933}].

\bibitem{Krisnanda} T. Krisnanda {\it et al}., ``Observable quantum entanglement due to gravity'', \href{https://www.nature.com/articles/s41534-020-0243-y}{npj. Quantum Inf. {\bf6}, 12 (2020)}.
\bibitem{Barton} G. Barton, ``Quantum mechanics of the inverted oscillator potential'', \href{https://www.sciencedirect.com/science/article/abs/pii/0003491686901429}{Annals Phys. {\bf166}, 322 (1986)}.

\bibitem{Yuce} C. Yuce, A. Kilic and A. Coruh, ``Inverted oscillator'', \href{https://iopscience.iop.org/article/10.1088/0031-8949/74/1/014/meta}{Phys. Scr. {\bf74}, 114 (2006)} [\href{https://arxiv.org/abs/quant-ph/0703234}{arXiv:quant-ph/0703234}].

\bibitem{Munoz} C.\,A. Mu\~noz, J.\, Rueda-Paz and K. Bernardo Wolf, ``Discrete repulsive oscillator wavefunctions'', \href{https://iopscience.iop.org/article/10.1088/1751-8113/42/48/485210/meta}{J. Phys. A: Math. Theor. {\bf42}, 485210 (2009)}.
\bibitem{Greiner} W. Greiner, {\it Quantum Mechanics: An Introduction}, 4th Edition
(Springer-Verlag, New York, 2001).

\bibitem{FormulasBook1} V.\,F. Zaitsev and A.\,D. Polyanin, {\it Handbook of Ordinary Differential Equations: Exact Solutions, Methods, and Problems}, 2nd Edition
(Chapman \& Hall/CRC Press, New York, 2002).

\bibitem{EngineeringBook} K.\,B. Wolf, {\it Integral Transforms in Science and Engineering} (Springer US, New York, 1979).

\bibitem{FormulasBook2} M.\, Abramowitz and I.\,A. Stegun, {\it Handbook of Mathematical Functions: with Formulas, Graphs, and Mathematical Tables} (Dover Publications, New York, 1965).

\bibitem{Olver} F.\,W.\,J. Olver, ``Uniform Asymptotic Expansions for Weber Parabolic Cylinder Functions of Large Orders", \href{https://www.nist.gov/nist-research-library/journal-research-volume-63b}{J. Res. Nat. Bur. Stds. B. Vol.\,{\bf63B}, 131 (1959)}.
\bibitem{Manasse} F.\,K. Manasse and C.\,W. Misner, and S.\,A. Werner, ``Fermi Normal Coordinates and Some Basic Concepts in Differential Geometry'', \href{https://aip.scitation.org/doi/10.1063/1.1724316}{J. Math. Phys. {\bf4}, 735 (1963)}.
\bibitem{Lambourne} R.\,J.\,A. Lambourne, {\it Relativity, Gravitation and Cosmology}, (Cambridge University Press, Cambridge, 2010).
\bibitem{Lewis1967} H.\,R. Lewis Jr., ``Classical and Quantum Systems with Time-Dependent Harmonic-Oscillator-Type Hamiltonians'', \href{https://journals.aps.org/prl/abstract/10.1103/PhysRevLett.18.510}{Phys. Rev. Lett. {\bf18}, 510 (1967)}.

\bibitem{Lewis1969} H.\,R. Lewis Jr. and W.\,B. Riesenfeld, ``An Exact Quantum Theory of the Time-Dependent Harmonic Oscillator and of a Charged Particle in a Time-Dependent Electromagnetic Field'', \href{https://aip.scitation.org/doi/abs/10.1063/1.1664991}{J. Math. Phys. {\bf10}, 1458 (1969)}.

\bibitem{WavePackets} D. Schuch, ``Riccati and Ermakov Equations in Time-Dependent and Time-Independent Quantum Systems'', \href{https://www.emis.de/journals/SIGMA/2008/043/}{SIGMA {\bf4}, 043 (2008)} [\href{https://arxiv.org/abs/0805.1687}{arXiv:0805.1687}].
\bibitem{GPE} F. Dalfovo {\it et al}., ``Theory of Bose-Einstein condensation in trapped gases'', \href{https://journals.aps.org/rmp/abstract/10.1103/RevModPhys.71.463}{Rev. Mod. Phys. {\bf71}, 463 (1999)} [\href{https://arxiv.org/abs/cond-mat/9806038}{arXiv:cond-mat/9806038}].
\bibitem{Anandan} J. Anandan, ``Gravitational and Inertial Effects in Quantum Fluids'', \href{https://journals.aps.org/prl/abstract/10.1103/PhysRevLett.47.463}{Phys. Rev. Lett. {\bf47}, 463 (1981)}; \href{https://journals.aps.org/prl/abstract/10.1103/PhysRevLett.52.401}{Erratum Phys. Rev. Lett. 52, 401 (1984)}.

\bibitem{Suarez} A. Su\'arez and P-H. Chavanis, ``Hydrodynamic representation of the Klein-Gordon-Einstein equations in the weak field limit: I. General formalism and perturbations analysis'', \href{https://journals.aps.org/prd/abstract/10.1103/PhysRevD.92.023510}{Phys. Rev. D {\bf92}, 023510 (2015)} [\href{https://arxiv.org/abs/1503.07437}{arXiv:1503.07437}].

\bibitem{KGIssue2} P-H. Chavanis and T. Matos, ``Covariant theory of Bose-Einstein condensates in curved spacetimes with electromagnetic interactions: the hydrodynamic approach'', \href{https://epjplus.epj.org/articles/epjplus/abs/2017/01/13360_2017_Article_1463/13360_2017_Article_1463.html}{Eur. Phys. J. Plus, {\bf132}, 30 (2017)} [\href{https://arxiv.org/abs/1606.07041}{arXiv:1606.07041}].
\bibitem{GPE1995A} M. Edwards and K. Burnett, ``Numerical solution of the nonlinear Schr\"odinger equation for small samples of trapped neutral atoms'', \href{https://journals.aps.org/pra/abstract/10.1103/PhysRevA.51.1382}{Phys. Rev. A {\bf51}, 1382 (1995)}.

\bibitem{GPE1995B} P.\,A. Ruprecht {\it et al}., ``Time-dependent solution of the nonlinear Schr\"odinger equation for Bose-condensed trapped neutral atoms'', \href{https://journals.aps.org/pra/abstract/10.1103/PhysRevA.51.4704}{Phys. Rev. A {\bf51}, 4704 (1995)}.

\bibitem{GPE1996A} F. Dalfovo and S. Stringari, ``Bosons in anisotropic traps: Ground state and vortices'', \href{https://journals.aps.org/pra/abstract/10.1103/PhysRevA.53.2477}{Phys. Rev. A {\bf53}, 2477 (1996)} [\href{https://arxiv.org/abs/cond-mat/9510142}{arXiv:cond-mat/9510142}].

\bibitem{GPE1996B} M. Edwards {\it et al}., ``Properties of a Bose-Einstein condensate in an anisotropic harmonic potential'', \href{https://journals.aps.org/pra/abstract/10.1103/PhysRevA.53.R1950}{Phys. Rev. A {\bf53}, R1950(R) (1996)}.

\bibitem{GPE1998} X-X. Yi, H-J. Wang and C-P. Sun, ``Bose-Einstein Condensation in Harmonic Oscillator Potentials'', \href{https://iopscience.iop.org/article/10.1088/0031-8949/57/3/002}{Phys. Scr. {\bf57}, 324 (1998)}.

\bibitem{GPEReview2013} X. Antoine, W. Bao and C. Besse, ``Computational methods for the dynamics of the nonlinear Schrodinger/Gross-Pitaevskii equations'', \href{https://www.sciencedirect.com/science/article/abs/pii/S0010465513002403?via%3Dihub}{Comput. Phys. Commun., {\bf184}, 2621, (2013)} [\href{https://arxiv.org/abs/1305.1093}{arXiv:1305.1093}].

\bibitem{R11} T. Damour, J.\,F. Donoghue, ``Equivalence Principle Violations and Couplings of a Light Dilaton'', \href{https://journals.aps.org/prd/abstract/10.1103/PhysRevD.82.084033}{Phys. Rev. D {\bf82}, 084033 (2010)} [\href{https://arxiv.org/abs/1007.2792}{arXiv:1007.2792}].

\bibitem{R12} T. Damour, ``Theoretical Aspects of the Equivalence Principle'', \href{https://iopscience.iop.org/article/10.1088/0264-9381/29/18/184001}{Class. Quantum Grav. {\bf29}, 184001 (2012)} [\href{https://arxiv.org/abs/1202.6311}{arXiv:1202.6311}].


\bibitem{Zamora-Zamora} R. Zamora-Zamora {\it et al}., ``Validity of Gross-Pitaevskii solutions of harmonically confined BEC gases in reduced dimensions'', \href{https://iopscience.iop.org/article/10.1088/2399-6528/ab360f}{J. Phys. Commun. {\bf3}, 085003 (2019)}.
\bibitem{Chandrasekhar} S. Chandrasekhar, {\it Ellipsoidal Figures of Equilibrium}, (Yale University Press, New Haven and London, 1969).

\bibitem{ConformalKG} F. Hammad, P. Sadeghi, N. Fleury and A. Leblanc, ``What can we learn from the conformal noninvariance of the Klein-Gordon equation?'', \href{https://arxiv.org/abs/2012.12355}{arXiv:2012.12355}.

\bibitem{Reality} H.\,R.\,Brown, {\it The Reality of the Wavefunction: Old Arguments and New.} In: Cordero A. (eds) Philosophers Look at Quantum Mechanics. \href{ https://doi.org/10.1007/978-3-030-15659-6_5}{Synthese Library (Studies in Epistemology, Logic, Methodology, and Philosophy of Science, vol 406. Springer, Cham. 2019)}.


\end{thebibliography}
\end{document}